\pdfoutput=1

\documentclass[aps,prd,reprint,twocolumn,showpacs,nofootinbib,superscriptaddress,floatfix]{revtex4-1}

\usepackage{amsmath,amssymb,amstext}
\usepackage[usenames,svgnames]{xcolor}
\usepackage{hyperref,graphicx}
\usepackage{subfigure}

\hypersetup{
    colorlinks=true,
    urlcolor=SteelBlue,
    linkcolor=red,
    citecolor=blue,
}

\renewcommand{\vec}[1]{\mathbf{#1}}

\begin{document}
\title{Unscreening scalarons with a black hole}

\author{Andrei V. Frolov}%
\email{frolov@sfu.ca}
\affiliation{Department of Physics, Simon Fraser University,\\
8888 University Drive, Burnaby, British Columbia V5A 1S6, Canada}

\author{Jos\'{e} T. G\'{a}lvez Ghersi}%
\email{joseg@sfu.ca}
\affiliation{Department of Physics, Simon Fraser University,\\
8888 University Drive, Burnaby, British Columbia V5A 1S6, Canada}
\date{\today}

\author{Alex Zucca}%
\email{azucca@sfu.ca}
\affiliation{Department of Physics, Simon Fraser University,\\
8888 University Drive, Burnaby, British Columbia V5A 1S6, Canada}
\date{\today}

\begin{abstract}
It is typically believed that the additional degrees of freedom in any modification of gravity are completely suppressed by the large energy densities coexisting with an astrophysical black hole. In this paper, we find that this might not always be the case. This belief holds for black holes formed via gravitational collapse in very dense environments, whereas the black holes with sufficiently low accretion rates that have low matter densities inside innermost stable circular orbit will generally unscreen chameleons. We develop a novel technique to study the dynamics of accretion of a scalar field onto a Schwarzschild-like black hole which is accurate on both short and long time scales. In particular, we study the behavior of the extra scalar degree of freedom in the Starobinsky and Hu-Sawicki $f(R)$ theories, for the symmetron model, and for the Ratra-Peebles model. Aside from calculating non-trivial static field profiles outside the black hole, we provide the tools to study the (in)stability and evolution towards the equilibrium solution for any generic well behaved set of parameters and initial conditions. Our code is made publicly available for further research and modifications to study other models.
\end{abstract}
\maketitle

\section{Introduction}

Einstein's theory of general relativity (GR) is one of the greatest achievements of modern science, due to its ability to describe many of the gravitational phenomena with impressive level of detail. Nevertheless, there are many motivations to look for alternative theories: the observation of cosmic acceleration, the unknown nature of dark matter and dark energy and the lack of an ultraviolet completion of general relativity related with the unavoidable presence of spacetime singularities in this theory. Any successful modification of Einstein's theory should solve some of these conundrums, while remaining consistent with the local tests of general relativity and do no harm to the standard behaviour of matter as discussed in Refs.~\cite{Will:2014kxa, Burgess:2013ara}. 

For example, $f(R)$ gravity was postulated in Ref.~\cite{Stelle:1976gc} as an attempt to produce a renormalisable theory of quantum gravity. It introduces a ghost-free functional of curvature, analogous to the notion of free energy in thermodynamics. Since its first incarnation, however, different forms of $f(R)$ have been used to solve issues at cosmological scales (see Refs.~\cite{Starobinsky:1980te, Hu:2007nk, Starobinsky:2007hu}). Notwithstanding, either considering a functional source of gravitational interaction different from the Ricci scalar, or simply adding extra gauge degrees of freedom in the gravitational sector has consequences at all energy scales. Gravity is not only defined by the metric, but also by other fields. The force exerted by any extra degree of freedom typically contradicts the stringent local constraints compatible with general relativity. Screening mechanisms such as theories in Refs.~\cite{Khoury:2003aq, Pietroni:2005pv, Vainshtein:1972sx} are designed to circumvent this issue by diluting the sources of this force.

In the case of an astrophysical black hole, one naively expects that every extra force will be screened by large environmental energy densities. In the particular case of a force sourced by a scalar, screening is expected to be a consequence of the no-hair theorem, as stated in Ref.~\cite{Bekenstein:1995un}, applicable in the case of a source with positive energy density. Nonetheless, as noticed in Ref.~\cite{Davis:2014tea}, the presence of a non-trivial distribution of accreting matter and the way this is coupled to the scalar field allows scalar hair outside the event horizon.

In this paper, we simulate the accretion of the additional scalar degree of freedom in the $f(R)$ models presented in Refs.~\cite{Hu:2007nk, Starobinsky:2007hu}, the symmetron model in Ref.~\cite{Hinterbichler:2010es} and the Ratra-Peebles chameleon in Ref.~\cite{Ratra:1987rm} in a dense environment. We assume that in such a region we can form a stable Schwarzschild black hole via gravitational collapse. This spherically symmetric black hole interacts with the screened extra real scalar field in a non-trivial matter distribution and accretes around the event horizon. The accreting matter density profile proposed here is a slight generalization to the suggested in Ref.~\cite{Davis:2014tea}. Even when the accretion in astrophysical black holes takes place in rotating spacetime solutions, it is interesting to explore the test case of a spherically symmetric system.

The purpose of this study is to visualize the process of accretion of the scalar field for various choices of model parameters, and discuss the time scales involved in the convergence to a non-trivial static profile. We designed a compact and efficient spectral code, flexible to modification and capable of producing time-dependent solutions of a scalar field in a Schwarzschild-like background. It is available for further application to other models at \url{https://github.com/andrei-v-frolov/accretion}.

The layout of this paper is as follows. In section \ref{sec:matterset}, we define the matter environment and the spacetime geometry in which the extra scalars propagate. In section \ref{sec:fr}, we review the equations of motion and particular features of the scalar degrees of freedom in the chosen $f(R)$ models and study how these fields accrete. In section \ref{sec:singularities}, we modify these models by adding terms proportional to $R^2$ and show the dynamical resolution of the curvature singularities in the Hu-Sawicki and the Starobinsky model. We extend our treatment for the symmetron model and the Ratra-Peebles chameleon in sections \ref{sec:symmetron} and \ref{sec:ratrapeebles}, respectively. Finally, we present summary of our results and conclusions in section \ref{sec:conclusions}. Numerical implementation is discussed in detail in Appendix~\ref{app:numerical}.

\section{Matter distribution for spherically symmetric black holes}
\label{sec:matterset}

Although astrophysical black holes are rotating systems, spherically symmetric solutions remain interesting when exploring some of the dynamical subtleties of the extra scalars in modified gravity. 
From Refs.~\cite{Scheel:1994yn, Frolov:2004rz}, we learn that the fate of a collapsing spherically symmetric system in standard scalar-tensor theories is to become a Schwarzschild black hole.

It is then sensible to ask whether the Schwarzschild solution remains a valid description of spacetime even in the presence of accreting matter. To answer this question, we briefly review the dynamics of a steady flow of matter to estimate the accretion rate of the black hole. 
We calculate the luminosity assuming the power lost by a generic inflow of particles traveling from infinity to the innermost stable circular orbit (ISCO) at $r_{\text{ISCO}}=6GM/c^2$
\begin{equation}
L=\left(1-\sqrt{\frac{8}{9}}\right)\dot{M}\sim 0.06\,\dot{M}.
\end{equation} 
Using the table of luminosities of active galactic nuclei included in Ref.~\cite{Khorunzhev:2012ga}, we note that the accretion rate of a supermassive black hole with $M_{\text{BH}}=10^9M_\odot$ is $\dot{M}\lesssim 1M_\odot/\text{yr}$. The accretion rate in a rotating black hole of similar luminosity could be even smaller, since $L\sim0.42\,\dot{M}c^2$ for near-extremal rotation. This implies that it would take $\sim 10^7$ years to change the mass of the black hole by 1\%. Therefore, it is reasonable to assume that spacetime is static within time scales we are interested in.

At this point of the discussion, it is necessary to provide an approximate expression of the matter density distribution outside the black hole's horizon. This is not by any means a full discussion of the radial structure equations for accretion disks. However, we provide sufficient arguments to justify our choice of a matter distribution. It must remain nearly static within the time scale estimated previously. For that purpose, we consider that radial matter density at a given radius $r$ is proportional to some positive power of the time that particles spend in orbits passing through that $r$. Stable orbits are possible only when $r\geq r_{\text{ISCO}}=6GM/c^2$. Therefore, the dynamics of a test particle moving in Schwarzschild spacetime only provides two scenarios in which matter can be found at $r< r_{\text{ISCO}}$: (I) these are on ``no-return'' trajectories towards the horizon or (II) particles are travelling on eccentric trajectories with a minimal radius smaller than $r_{\text{ISCO}}$. The latter case is highly unlikely as viscous forces spread anisotropies all along the accretion disk via diffusion. In Ref.~\cite{Frank:2002bk}, one can find a simplified linear model explaining that the cause of viscosity is the radial propagation of angular momentum from one orbit to another in the disk of matter. In this case, a final state of radial homogeneity is reached very rapidly. In a more realistic scenario, diffusion is driven by non-linear viscous forces studied in magnetohydrodynamical (MHD) simulations of rotating (and slightly magnetized) systems, as explained in Ref.~\cite{Abramowicz:2011xu}. In any case, the probability of finding matter on orbits within innermost stable circular orbit (ISCO) is greatly reduced, and in consequence, so is the matter density in those regions. For black holes with low accretion rates, this density drop could be sufficient to unscreen the scalarons within immediate vicinity of the black hole, as we will show in the following sections.

Using the arguments aforementioned and considering $\sigma \in [0,1)$ as the density contrast parameter, our rough prescription for the pressureless matter distribution is 
\begin{equation}
\rho= 
     \begin{cases}
       \sigma\rho_0 ;&\quad r_g<r<r_{\text{ISCO}}\\
       \rho_0; &\quad r\ge r_{\text{ISCO}},
     \end{cases}
\end{equation}
where $r_g=2GM/c^2$ is gravitational radius where black hole horizon is located. This coincides with the distribution suggested in Ref.~\cite{Davis:2014tea} when $\sigma=0$. For numerical reasons, the use of a smooth matter profile approximated by a hyperbolic tangent is more convenient
\begin{equation}
\label{eqn:rho}
\rho=(1-\sigma)\rho_0\frac{1+\tanh a_0(r-r_{\text{ISCO}})}{2} + \sigma\rho_0,
\end{equation}
where $a_0\gg 1$ and $\rho_0> 0$. The shape of this matter distribution is a crude approximation of the results presented in Ref.~\cite{Abramowicz:2011xu} when $\sigma < 10^{-4}$. As a consequence, we will setup all the wave equations for the scalar field using the Schwarzschild geometry 
\begin{equation}
ds^2=-\left(1-\frac{r_g}{r}\right)dt^2+\frac{dr^2}{1-\frac{r_g}{r}}+r^2d\Omega^2.
\end{equation}
and the matter density profile suggested in \eqref{eqn:rho}. However, it is convenient to change from Schwarzschild to tortoise coordinates $dr=\left(1-r_g/r\right)dx$
\begin{equation}
\label{eqn:Schwarzschildtort}
ds^2=\left(1-\frac{r_g}{r(x)}\right)\left[-dt^2+dx^2\right]+r^2(x)d\Omega^2.
\end{equation}
The coordinate change from $r$ to $x$ and its inversion is discussed in detail in Appendix \ref{app:numerical:tortoise}. We neglect back-reaction of the scalaron dynamics on the background geometry for all the scalar-tensor theories explored in this project, and treat the metric as static in the scalaron equations of motion. Using \eqref{eqn:Schwarzschildtort}, we can write the equation of motion of the scalar field $\Box\phi=V_{\text{eff}}'(\phi)$ which appears in \eqref{eqn:scalareqmov} and \eqref{eqn:symmetron} as a spatially damped wave equation in 1+1 dimensions
\begin{equation}
\label{eqn:eqtosolve}
\begin{split}
\biggl[-\frac{\partial^2}{\partial t^2}+ & \frac{\partial^2}{\partial x^2}+\frac{2}{r(x)}\left(1-\frac{r_g}{r(x)}\right)\frac{\partial}{\partial x}\biggr]\phi(t,x) \\
& -\left(1-\frac{r_g}{r(x)}\right)V^{\prime}_{\text{eff}}(\phi)=0.
\end{split}
\end{equation}
This choice of coordinates is sufficient for our purposes since the scalar solutions we seek do not need to cover the black hole's interior $(r<r_g)$. For simplicity of the implementation, we use units of $r_g=1$ in the code, but other choices can also be considered without difficulty.

\section{Scalar accretion in $f(R)$ theories}
\label{sec:fr}
In this section we describe the two examples of $f(R)$ theories considered in this work: the Starobinsky and Hu-Sawicki models, where the Ricci scalar $R$ in the Einstein-Hilbert action is replaced by a function of $R$. We will consider equations of motion derived in the metric formalism. Alternative formalisms like the Palatini or the metric-affine mentioned in Refs.~\cite{Sotiriou:2008rp, Olmo:2005hd, Sotiriou:2006qn, BeltranJimenez:2017doy} can change the number and/or the nature of the degrees of freedom that emerge. In particular, in Ref.~\cite{Sotiriou:2008rp}, we see that the Palatini formulation of $f(R)$ is equivalent to a $\omega_0=-3/2$ Brans-Dicke theory, and no new dynamical degrees of freedom appear. In all cases, one needs to ensure that the model remains ghost-free in the gravity sector and that there are no tachyonic modes for it to be viable. A recent discussion on ghosts in various formulations appeared in Ref.~\cite{Koivisto:2013kwa}.

\subsection{Chameleons in Starobinsky and Hu-Sawicki model}
\label{sec:fR}
\begin{figure*}
\centering
\subfigure{
\includegraphics[width=.47\textwidth]{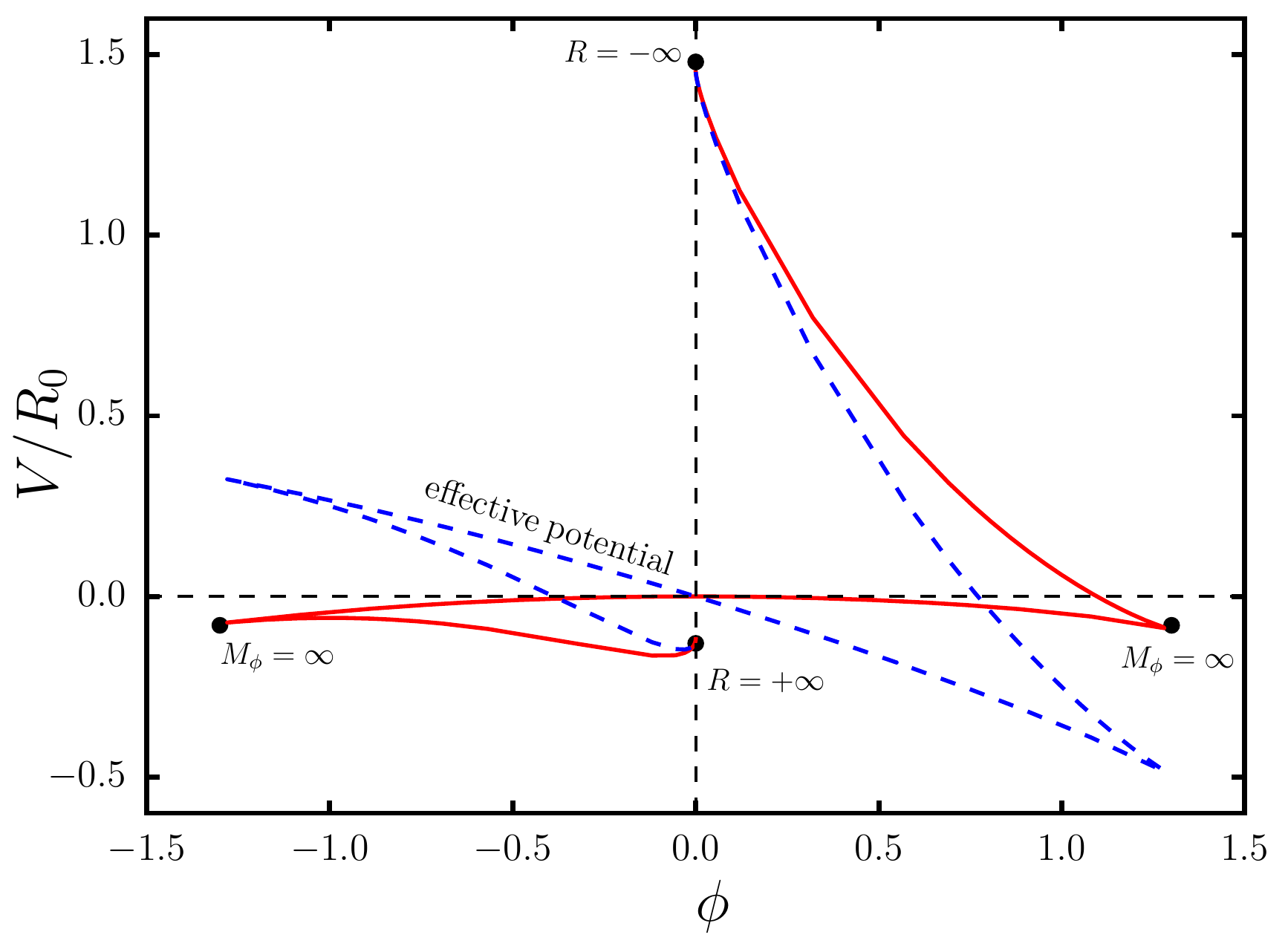}} \,
\subfigure{
\includegraphics[width=.47\textwidth]{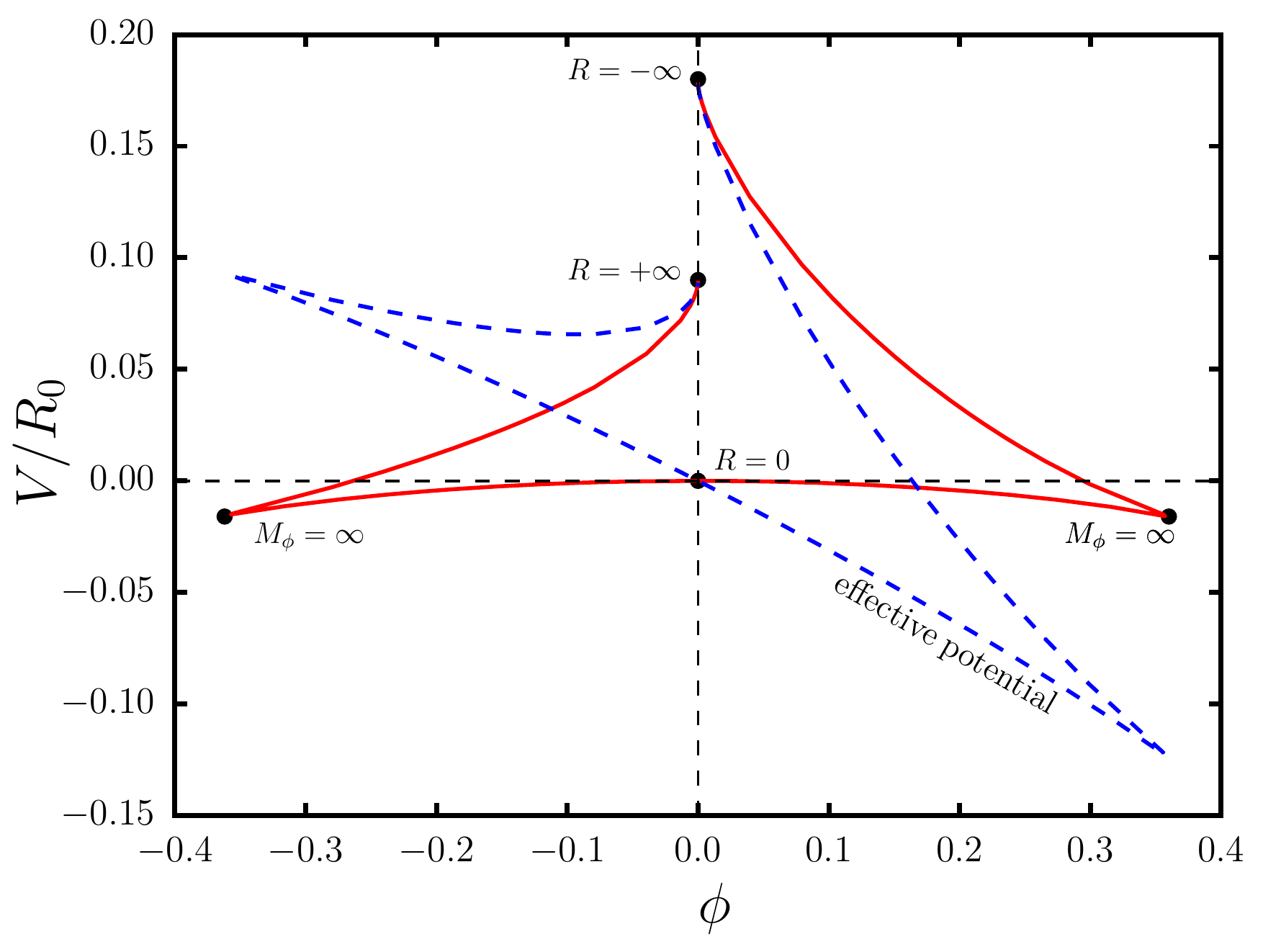}}
\caption{\label{fig:potentials} Left panel: bare and effective potential for $\lambda=2.0$ and $n=1$ in the Starobinsky model. Matter defines an effective minimum and a mass for the field. Right panel: bare and effective potential for $\alpha=0.8$ and $n=2$ in the Hu-Sawicki model. Increasing densities of surrounding matter also define a steady-state solution. Both effective potentials are concave up around the equilibrium position, which means that the scalarons are not tachyonic.} 
\end{figure*}

We first consider $f(R)$ theories described, in the Jordan frame, by the action
\begin{equation}
\label{eqn:action}
S=\int \frac{f(R)}{16\pi G}\,\sqrt{-g}\,d^4x + S_m[g_{\mu\nu},\psi],
\end{equation}
where $g_{\mu \nu}$ is the Jordan frame metric and $\psi$ are the matter fields. Varying the action with respect to the metric $g_{\mu\nu}$ we obtain equations of motion which replace Einstein's equation in the $f(R)$ models; they are
\begin{equation}
\label{eqn:eqmovtensor}
\begin{split}
f_R R_{\mu\nu} - \frac{1}{2}f&g_{\mu\nu} = 8\pi G\, T_{\mu\nu}+\\
&\nabla_{\mu} \nabla_{\nu} f_R-g_{\mu\nu}\Box f_R,
\end{split}
\end{equation}
where $f_R\equiv \partial f/\partial R$. The two terms in the second line of the equation \eqref{eqn:eqmovtensor} contain fourth-order derivatives of the metric, a signal that a new degree of freedom emerges in the theory. This can be seen explicitly by taking the trace of the equation above
\begin{equation}
\label{eqn:fReom}
\Box f_R = \frac{1}{3}\,(2f - f_R R) + \frac{8\pi G}{3}\, T, 
\end{equation}
which yields a second order equation of motion for the real field $f_R$ with a canonical kinetic term under the influence of an effective potential $V^{\prime} \equiv (2f-f_R R)/3$ and an external force term ${\cal F} \equiv -8\pi G\, T/3$ with $T \equiv T^{\mu}_{\mu}$. By defining $\phi \equiv f_R - 1$ we can rewrite equation \eqref{eqn:fReom} simply as
\begin{equation}
\label{eqn:scalareqmov}
\Box \phi = V^{\prime}(\phi) - {\cal F},
\end{equation}
where prime denotes derivative with respect to $\phi$. Alternatively, one can explicitly see the emergence of the extra degree of freedom, usually dubbed ``scalaron'', by mapping the action \eqref{eqn:action} into the Einstein frame, as described in Ref.~\cite{Maeda:1988ab, DeFelice:2010aj}.  Solving the equation \eqref{eqn:scalareqmov} requires the knowledge of the potential $V(\phi)$ which is defined in a parametric form via
\begin{equation}
\frac{dV}{dR}=\frac{dV}{d\phi}\frac{d\phi}{dR}=\frac{1}{3}\left(2f - f_R R\right) f_{RR},
\end{equation} 
or, integrating with a choice $V\left|_{R=0}\right.=0$ for a constant,
\begin{eqnarray}
\phi(R) &=& f_R-1,\\
V(R) &=& \frac{1}{3}\int\limits_0^R d\tilde{R}\left(2f(\tilde{R}) - f'(\tilde{R}) \tilde{R} \right) f''(\tilde{R}). \label{eqn:potential}
\end{eqnarray}
When plotting the scalar potentials in Fig.~\ref{fig:potentials}, we observe that they are generally multi-valued, with turning points at field values where $f''(R)=0$. One must be aware of the branch choice when determining the curvature value $R$ and the effective force $V'(\phi)$ for the field $\phi$ in the wave equation \eqref{eqn:scalareqmov}. The branch we are interested in is the one connected to the large curvature $R \rightarrow +\infty$ where Einstein gravity is recovered by screening. The particular models we study are defined by specific forms of $f(R)$
\begin{gather}
\label{eqn:fstarobinsky}
f_{\text{S}}=\displaystyle{R+\lambda\left[\frac{1}{\left(1+(R/R_0)^2\right)^n}-1\right]R_0},\\
\label{eqn:fhusawicki}
f_{\text{HS}}=\displaystyle{R-\frac{\alpha(R/R_0)^nR_0}{1+\beta(R/R_0)^n}},
\end{gather}
which correspond to the Starobinsky -- in Ref.~\cite{Starobinsky:2007hu} -- and Hu-Sawicki -- in Ref.~\cite{Hu:2007nk} -- models respectively. Hence, by replacing the solution of \eqref{eqn:potential} in $f_{\text{S}}$ or $f_{\text{HS}}$ (and their derivatives) the potentials are completely determined as functions of $\phi$ for every physical choice of parameters. A particular feature of \eqref{eqn:fhusawicki} is that it has an apparent extra parameter compared to \eqref{eqn:fstarobinsky}. However, it is entirely free, and we can reduce the number of parameters by considering the transformation $\alpha\rightarrow \lambda^a\alpha$, $\beta\rightarrow \lambda^b\beta$ and $R_0\rightarrow \lambda^cR_0$. $f_{\mathrm{HS}}$ is invariant under this transformation if 
\begin{eqnarray}
\label{eqn:invariance}
&a-(n-1)c=0,\\
&b-nc=0,\nonumber
\end{eqnarray}
for a given value of $n$. It is therefore possible to map any solution for one set of parameters to an equivalent one for a different set following \eqref{eqn:invariance}, hence we fixed the parameter $\beta$ to be 1 throughout the rest of the paper and one can convert to other choices via these transformations. Our choices of $\beta$ and the crossover curvature scale $R_0$ are useful to compare these results with the solutions from the Starobinsky model. As summarized in Ref.~\cite{Sotiriou:2008rp}, $f(R)$ models of gravity in metric formalism must have $f'(R)>0$ and $f''(R)>0$ to avoid ghost-like gravitons and tachyonic scalarons, respectively.

From equation \eqref{eqn:scalareqmov}, we can define an environmentally dependent effective potential $V_{\text{eff}}(\phi)$ that provides the same equations of motion in the regions of constant force ${\cal F}$ by
\begin{equation}
\label{eqn:veff}
V_{\text{eff}}(\phi)=V(\phi) - {\cal F} \phi.
\end{equation}
One peculiar feature of the effective potential is that the extra term coming from the interaction with matter provides an external source term. As a result of this, the no-hair theorem in its usual form is in general not applicable. In this paper, the presence of matter with $T^{\mu}_{\mu}\neq 0$ (which excludes electromagnetic radiation) is not neglected.

One must consider any particular choice of model parameters for $f_\text{HS}$ and $f_\text{S}$ that could emerge from their corresponding renormalization group flows. Therefore, it is prudent to study the flow lines in parameter space by exploring the stability of the scalar wave equation for different choices of model parameters, even when we consider cases where the model does not match with current observations. The field solutions are screened in the same way as we described in section \ref{sec:ratrapeebles}, henceforth $f_{\text{S}}$ and $f_{\text{HS}}$ scalarons can also be dubbed as ``chameleons''. 

\subsection{Accreting chameleons in Hu-Sawicki and Starobinsky models}

\begin{figure*}
\centering
\subfigure{\includegraphics[width=0.42\textwidth]{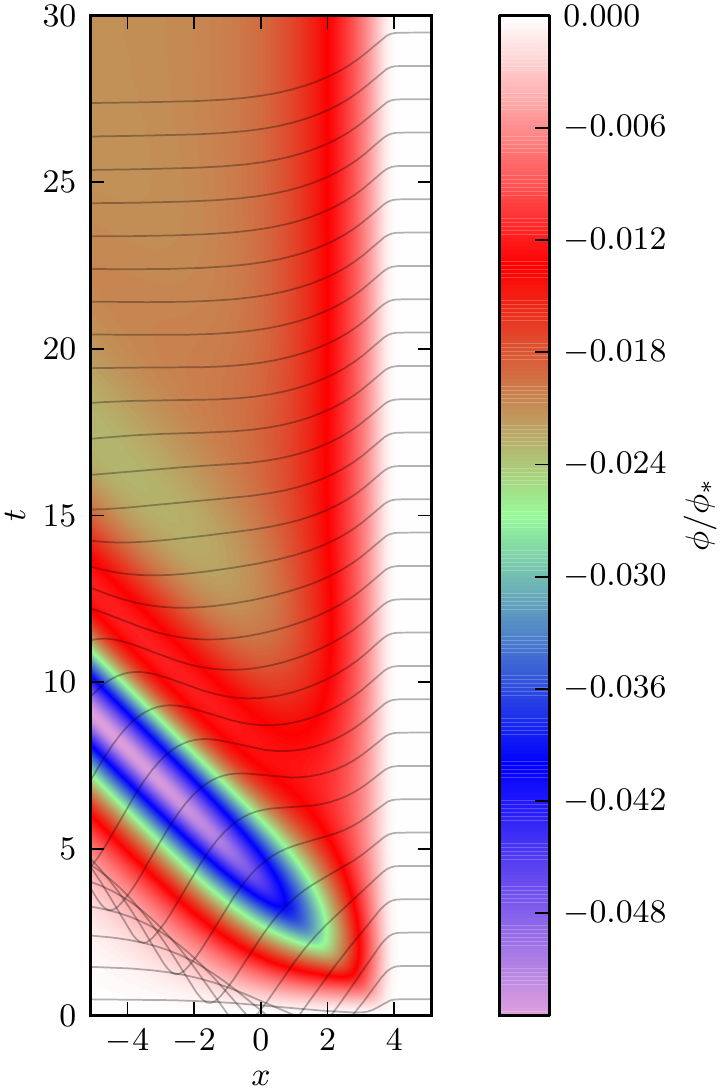}} \,
\subfigure{\includegraphics[width=0.42\textwidth]{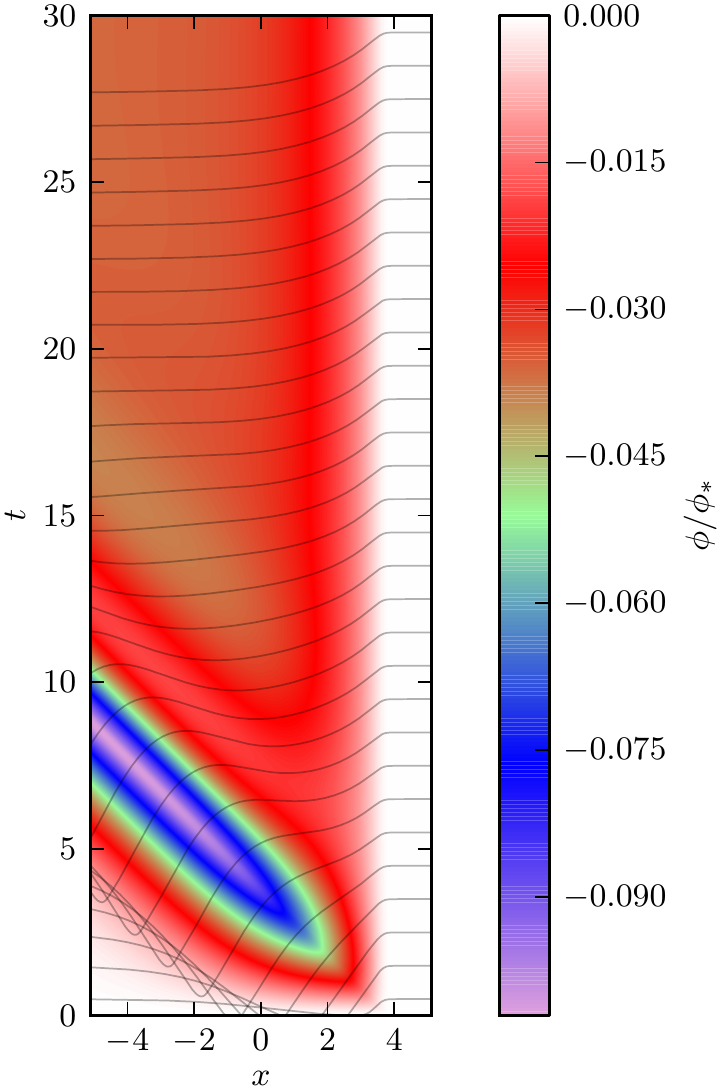}}
\caption{\label{fig:evolution} Accretion of the field in the Starobinsky (left panel) and Hu-Sawicki (right panel) models in tortoise coordinates. Field profiles are plotted on top of the intensity map: each line represents the amplitude of the field at constant time. The values of the field are consistent with the condition $f'(R)>0$.}
\end{figure*}

Now we present our results after evaluating the scalar equations of motion in \eqref{eqn:scalareqmov}. In order to proceed, we must first find the effective potentials corresponding to \eqref{eqn:fstarobinsky} with arbitrary densities of matter.


In the left panel of Fig.~\ref{fig:potentials} we observe that the features of this potential correspond to what is usually called a chameleon field. Just as described in Ref.~\cite{Brax:2008hh}, the steepness and depth of the potential grows with the surrounding matter density in which the field propagates. The effective potential of the Hu-Sawicki model is plotted in the right panel of Fig.~\ref{fig:potentials}.


In both cases, we observe that the formation of curvature singularities do not require infinite energies to be achieved. In addition to that, we notice the existence of an equilibrium configuration for the field, the corresponding minimum in the effective potential $V_{\text{eff}}'(\phi) = 0$ is defined by
\begin{equation}
V'(\phi)={\cal F}.\label{equilibrium}
\end{equation} 
In the presence of accreting matter outside $r_{\text{ISCO}}=3\, r_g$ the field is screened, which implies a very small value of it in this region. Inside $r_{\text{ISCO}}$, we can set the matter density in \eqref{eqn:rho} to zero for now. Naively, one should not expect significant differences in our results calculated in the Jordan frame when compared with what is expected in the Einstein frame after conformal transformations: inside $r_{\text{ISCO}}$, the matter density is zero regardless of any value of the field; outside, the screening sets the conformal coupling to one. It is typically argued that there are changes in the metric and equations of motion of a test particle when these entities are rephrased in this frame: the derivatives of the conformal coupling might modify these entities in a non-negligible way. This is true in general, however these changes do not represent a significant contribution to the solutions we present since these modifications are always proportional to the first and second radial derivatives of the field, which is a smooth function outside the event horizon.

In order to find the screening value of $\rho_0$ in \eqref{eqn:rho}, we evaluate the equilibrium condition in (\ref{equilibrium}) at the screened value of the field $\phi_0$ far away from $r_{\text{ISCO}}$ (which is very close to zero). Numerically, this is more convenient than (but still equivalent to) finding the equilibrium value of $\phi$ for a given value of $\rho_0$ from (\ref{equilibrium}) since the dependence of $V'(\phi)$ on the field is implicit.

The evolution of the system requires initial conditions. In what follows, we set the units of the field $\phi$ in terms of a pivot value $\phi_*$, which for $f(R)$ models is dimensionless and is chosen to be simply $\phi_* = 1$. We consider a completely screened initial configuration of the field $\phi(r,t_0) = -10^{-4}\phi_*$ which is originally spatially homogeneous, and assume $\dot{\phi}(r,t_0)=0$ for the initial field velocity. Our choice of initial conditions is the same in our treatment of both $f(R)$ models we present here. In the left panel of Fig.~\ref{fig:evolution}, we see the accretion of the Starobinsky chameleon until it approaches to its static solution, choosing $n=1$, $\lambda=2.0$ and $R_0=10^{-2}/r_g^2$ as the model parameters to run the simulation.


The chameleon accretes around the horizon, then it oscillates slightly around the static solution. Gradients do not cancel outside the horizon for $r\leq r_{\text{ISCO}}$, which make the ``hair'' profile non-trivial. In addition to this, the screening outside $r_{\text{ISCO}}$ is not lost during accretion. Which shows that the equilibrium condition in (\ref{equilibrium}) is being held.

Field evolution of the Hu-Sawicki chameleon is plotted in the right panel of Fig.~\ref{fig:evolution}, where we used the the same initial conditions. $n=1$, $\alpha=2.0$ and $R_0=10^{-2}/r_g^2$ were the model parameters chosen for the numerical evolution. Hu-Sawicki chameleon also accretes around the event horizon resulting in a non-trivial hair solution. We study the negative field branches of the potentials in Fig.~\ref{fig:potentials} that roughly scale as $\phi^{1/\kappa}$ with $\kappa >2$, therefore the field becomes less massive in the regions where its amplitude deviates from the screened value. 

\begin{figure}
\begin{center}
\includegraphics[width=0.42\textwidth]{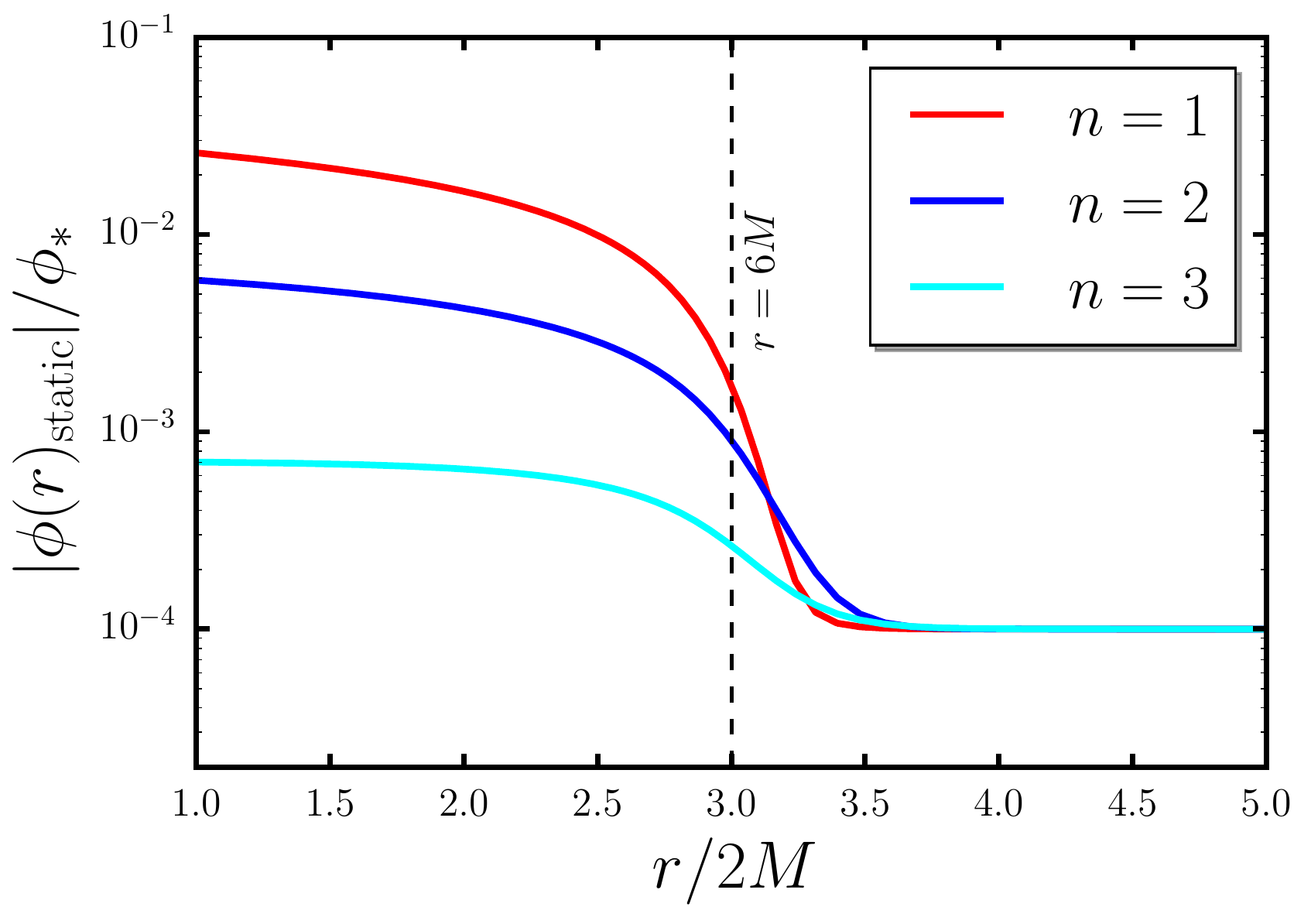}
\caption{Static profiles of the scalar field in the Starobinsky model for $\lambda=1.0$ and varying $n$.}
\label{fig:starns}
\end{center}
\end{figure}

\begin{figure}
\begin{center}
\includegraphics[width=0.42\textwidth]{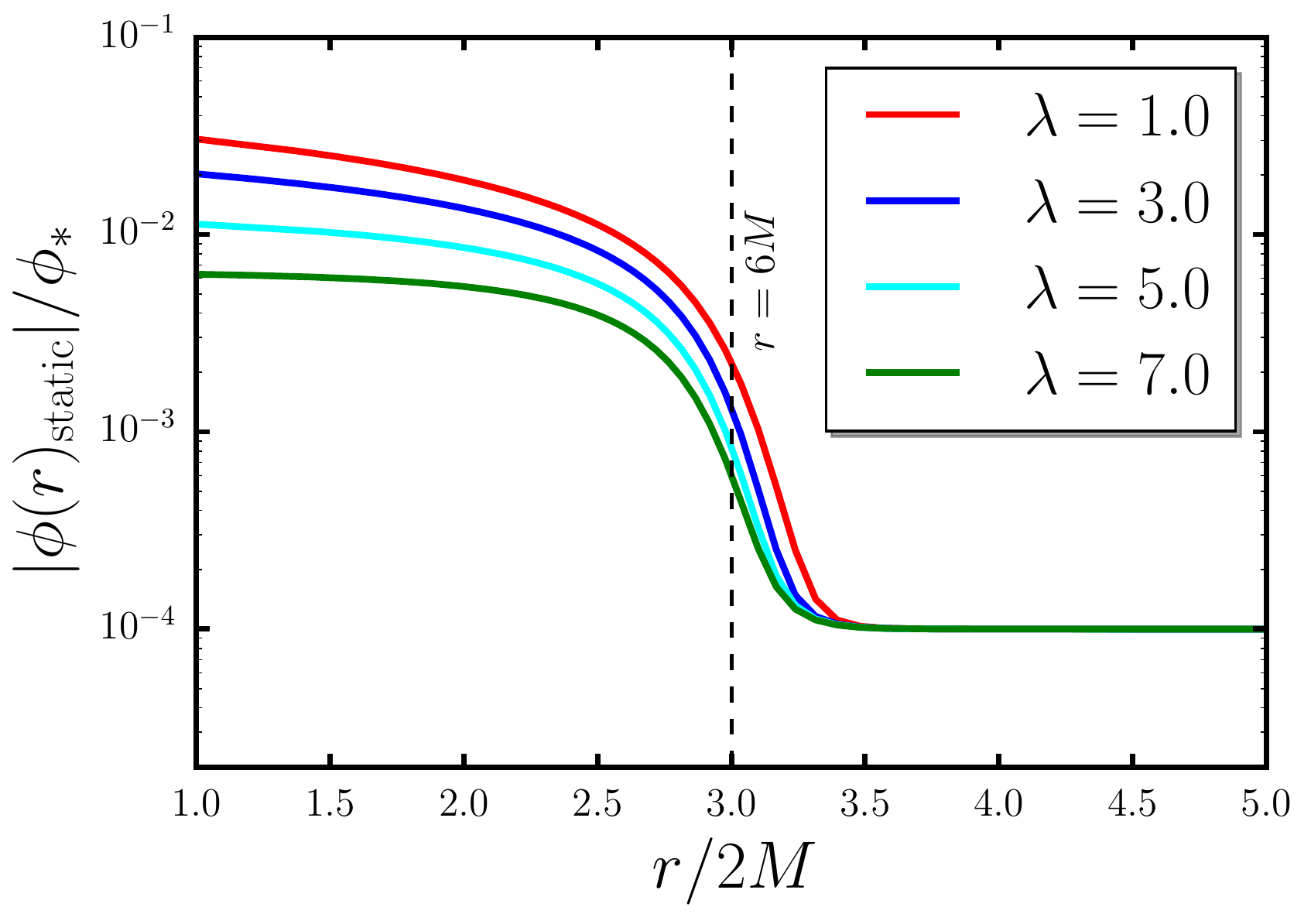}
\caption{Static profiles of the scalar field in the Starobinsky model for $n=1$ and varying $\lambda$.}
\label{fig:starlambdas}
\end{center}
\end{figure}

The convergence into a hair solution as well as its shape are sensitive to choice of model parameters. Static solutions are found using the relaxation method described in Appendix \ref{app:numerical:static}, where we also discuss all the details related to the numerical evolution. In Fig.~\ref{fig:starns}, we evaluate the solutions of \eqref{eqn:eqtosolve} in the static limit for different parameters of the Starobinsky model. We find the solutions for different values of $n$, while keeping $\lambda=1.0$ as a constant. Likewise, in Fig.~\ref{fig:starlambdas} we evaluate the change of the static field profile when $\lambda$ varies and $n=1$ is kept as a constant. In the same way, we represent static solutions for different values of the density contrast parameter $\sigma$.

Different values of $\lambda$ and $R_0$ define how effective is the modification of gravity with respect to GR. In particular, $\lambda$ controls the depth and vertical extension of the effective potential and in consequence, it affects the existence and stability of the field solutions, while $R_0$ sets the crossover curvature scale. In particular, we chose $R_0=10^{-2}/r_g^2$ to have the same value throughout this paper.   

\begin{figure}
\begin{center}
\includegraphics[width=0.42\textwidth]{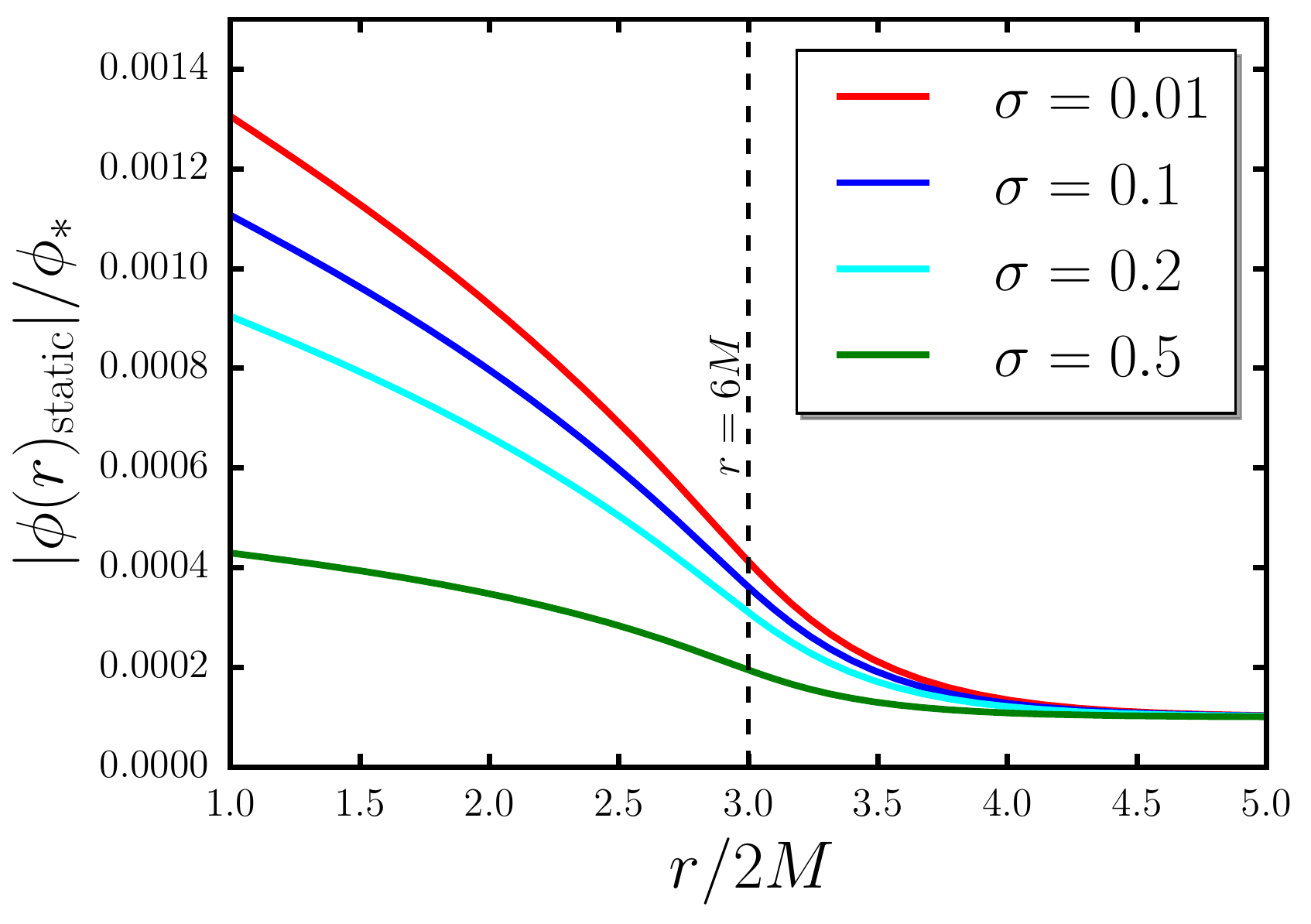}
\caption{Static profiles of the scalar field in the Starobinsky model for $n=1$ and varying density contrast $\sigma$.}
\label{fig:starsigmas}
\end{center}
\end{figure}

\begin{figure}
\begin{center}
\includegraphics[width=0.42\textwidth]{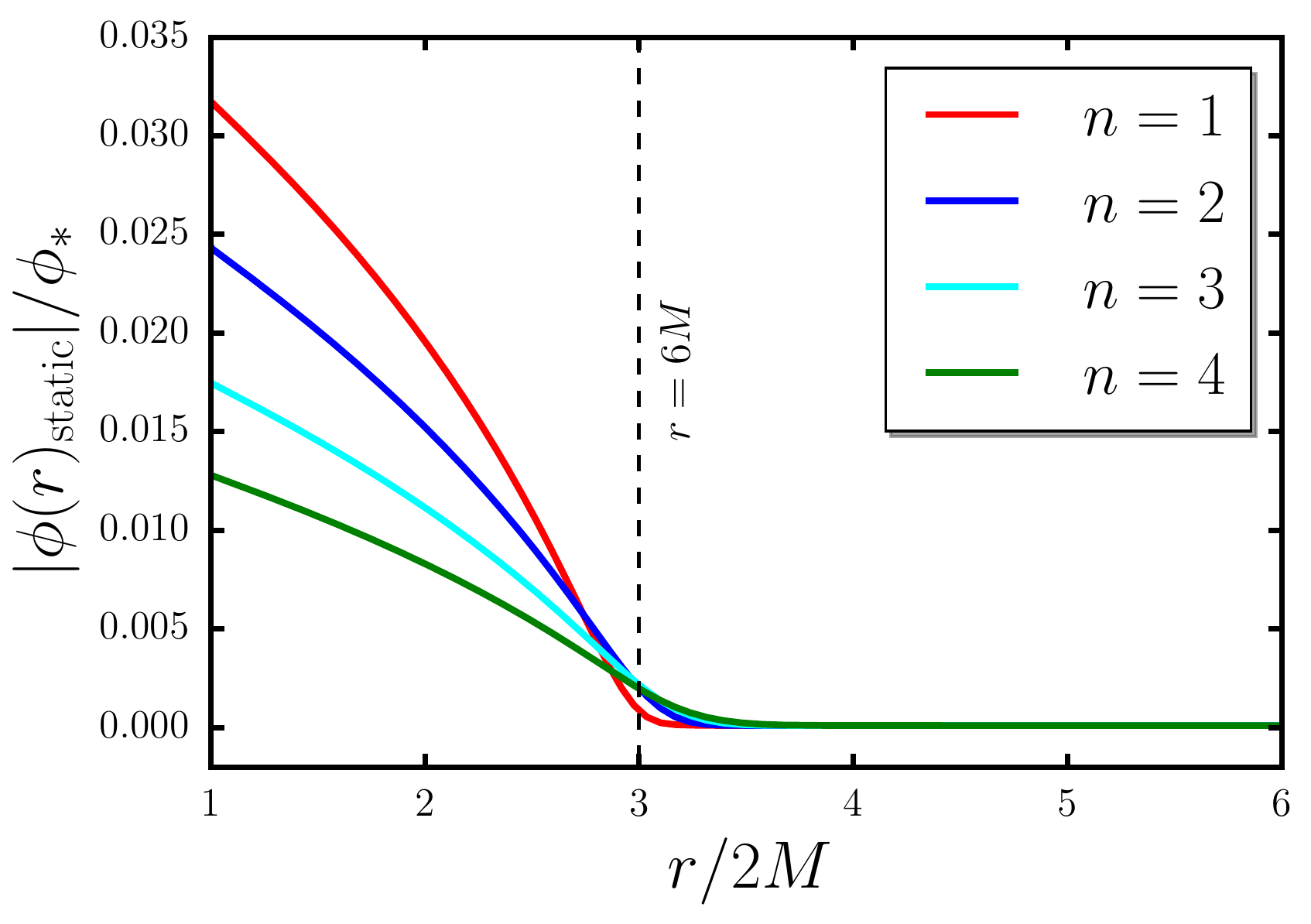}
\caption{Static profiles of the scalar field in the Hu-Sawicki model for $\alpha=1.0$ and varying $n$.}
\label{fig:hsns}
\end{center}
\end{figure}

\begin{figure}
\begin{center}
\includegraphics[width=0.42\textwidth]{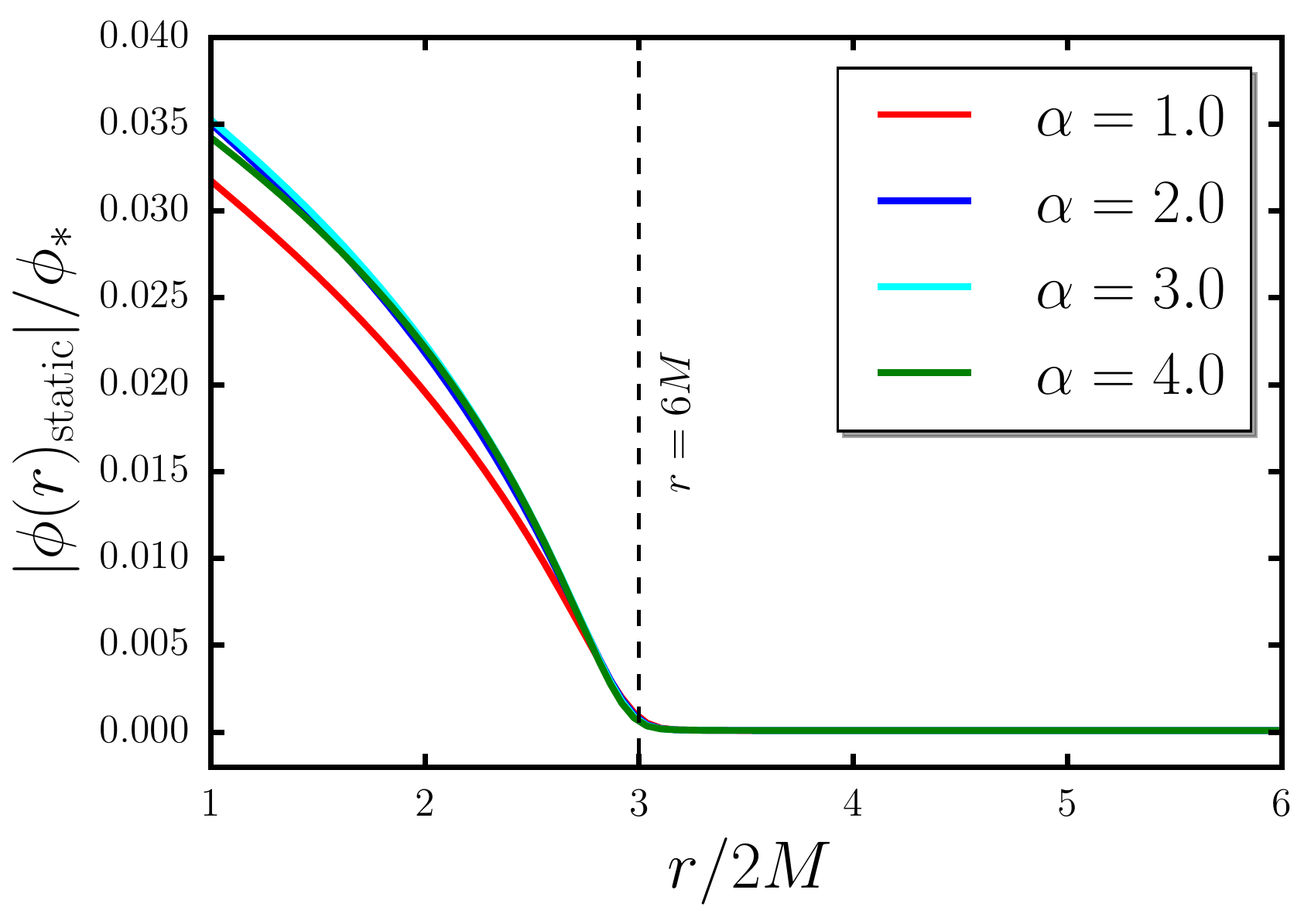}
\caption{Static profiles of the scalar field in the Hu-Sawicki model for $n=1$ and varying $\alpha$.}
\label{fig:hsalphas}
\end{center}
\end{figure}

\begin{figure}
\begin{center}
\includegraphics[width=0.42\textwidth]{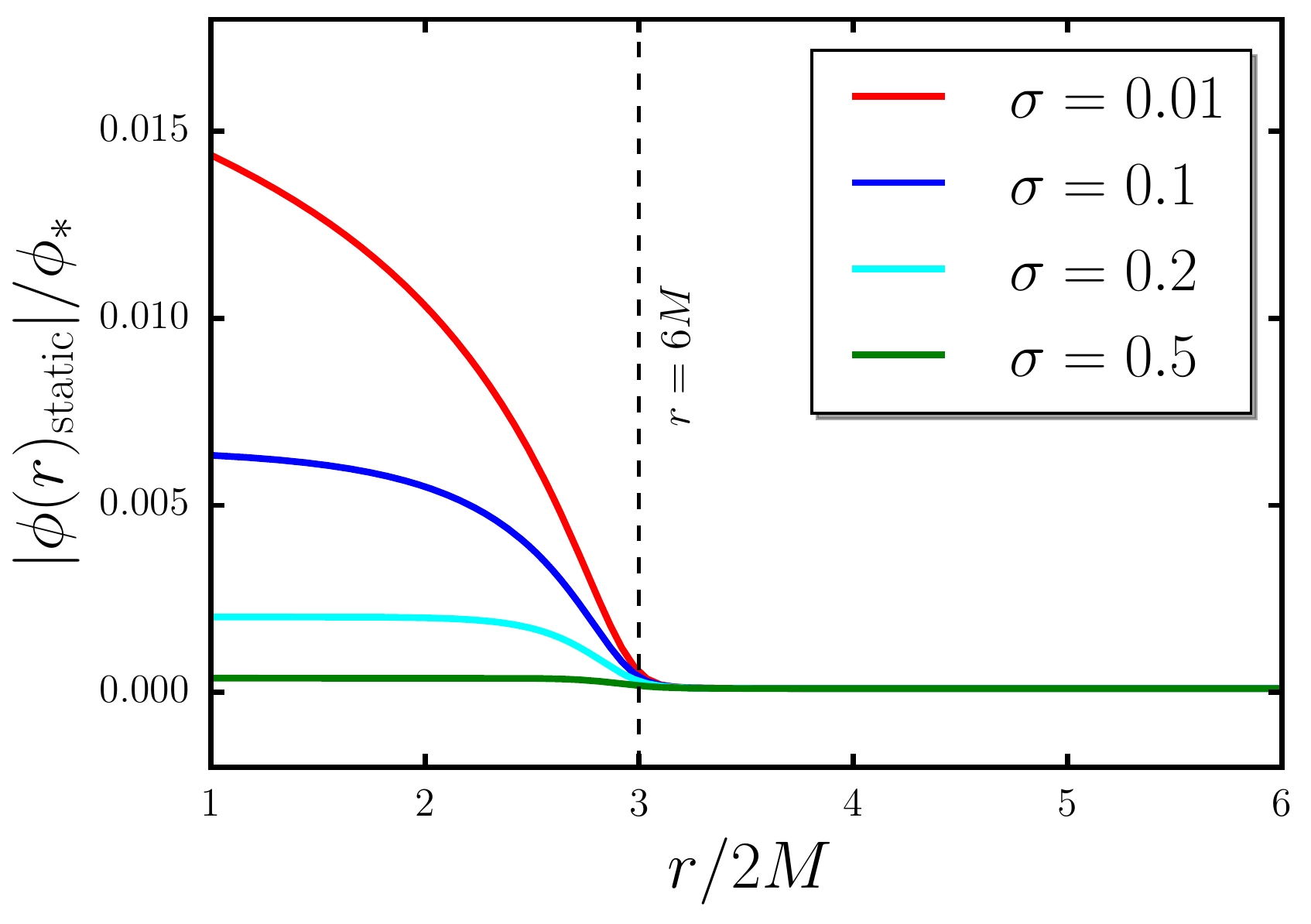}
\caption{Static profiles of the scalar field in the Hu-Sawicki model for $n=1$ and varying density contrast $\sigma$.}
\label{fig:hssigmas}
\end{center}
\end{figure}

\begin{figure*}
\centering
\subfigure{
\includegraphics[width=0.43\textwidth]{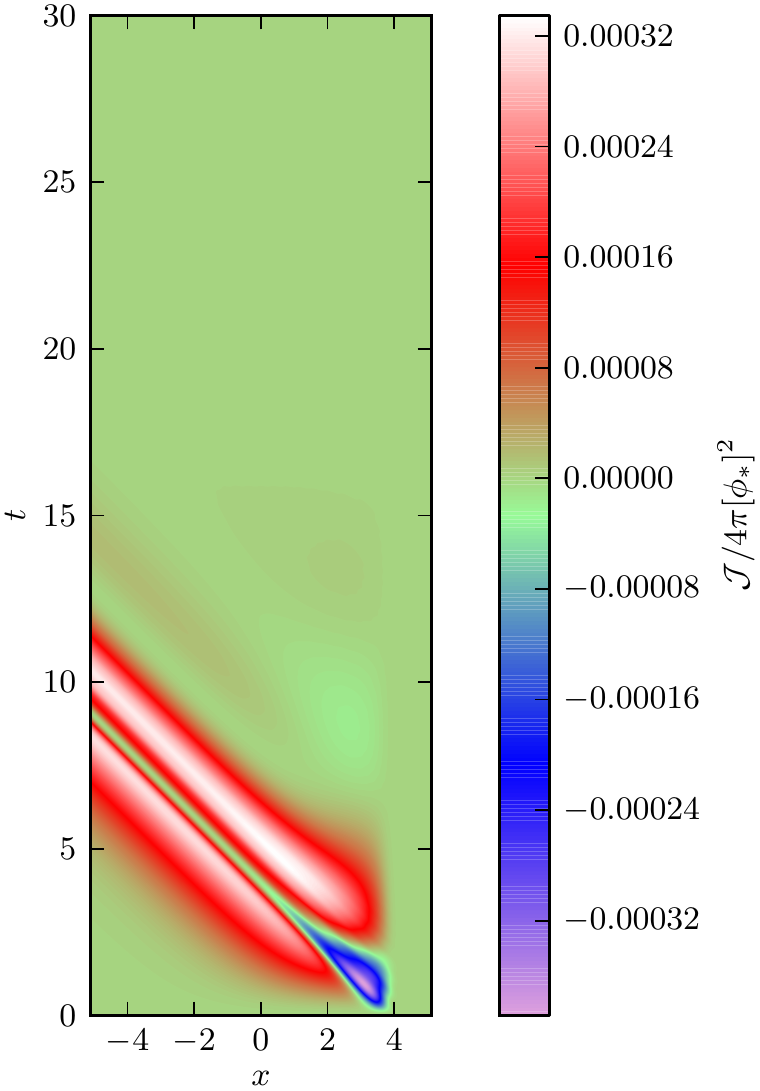}} \,
\subfigure{
\includegraphics[width=0.42\textwidth]{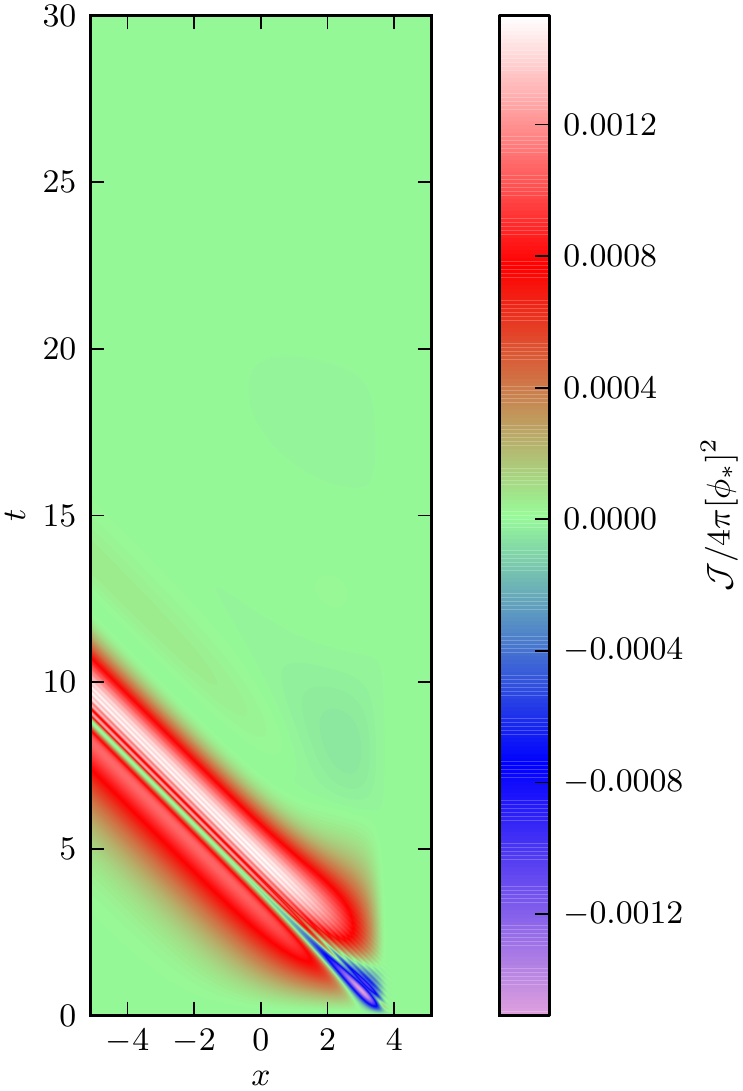}}
\caption{\label{fig:fluxes} Scalar field flux accreting towards horizon in the Starobinsky model (left panel) and in the Hu-Sawicki model (right panel).}
\end{figure*}

As expected, we notice the reduction of the field amplitude in $r\in (r_g, r_{\text{ISCO}}]$ for larger values of $\sigma$. However, it is not required to impose $\sigma=0$ to obtain a hairy solution. Furthermore, we calculate the static solutions corresponding to different choices parameters of the Hu-Sawicki model. The shape of the static solution for different values of $n$ can be found in Fig.~\ref{fig:hsns}. Additionally, steady-state solutions for different values of $\alpha$ and $\sigma$ are represented in Fig.~\ref{fig:hsalphas} and Fig.~\ref{fig:hssigmas}, respectively. In all of these cases, the rest of the parameters were kept as constants. The dynamical results exposed show the accretion of the chameleon solutions and proof the existence of non-trivial stable solutions of the static version of \eqref{eqn:eqtosolve} outside $r=r_g$ for the parameters we chose. The radial scalar flux ${\cal J} \equiv 4\pi r^2\, T[\phi]^r_0=4\pi r^2\,\phi_{, x}\phi_{,t}$ is represented in Fig.~\ref{fig:fluxes}, showing no propagation outside $r_{\text{ISCO}}$.

\begin{figure}
\begin{center}
\includegraphics[width=0.42\textwidth]{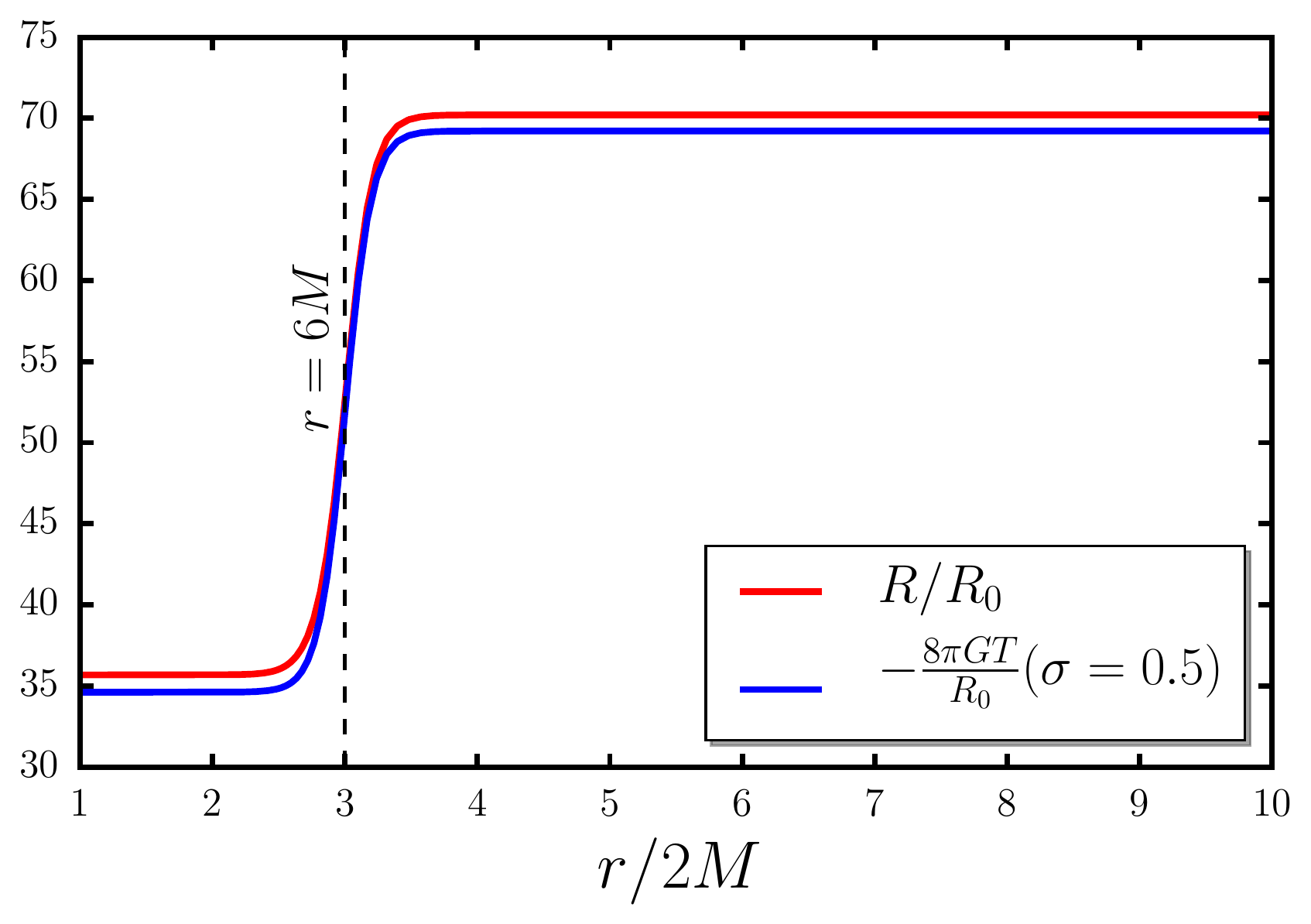}
\caption{Checking for local scalaron equilibrium in fully screened configuration with $\sigma=0.5$ in the Hu-Sawicki model. This result is very similar in the case of the Starobinsky model.}
\label{fig:equilHS}
\end{center}

\begin{center}
\includegraphics[width=0.42\textwidth]{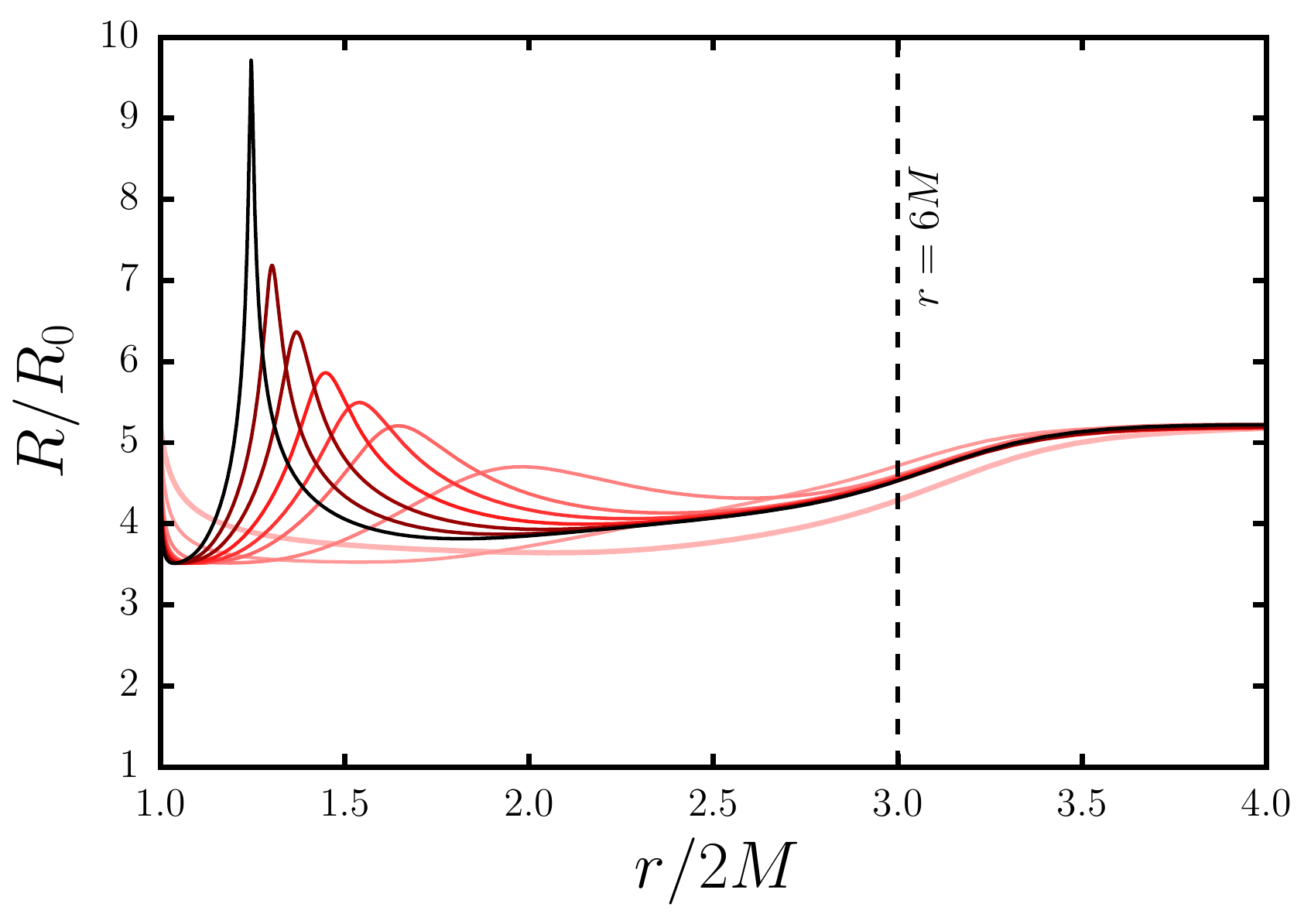}
\caption{Formation of a naked singularity for $n=3$ and $\lambda=2.1$ in the Starobinsky model.}
\label{fig:nakedST}
\end{center}
\end{figure}

We show an additional way to test that the static profiles in Figs.~\ref{fig:starns}-\ref{fig:hsalphas} are suitable representations of field configurations around the minimum of the effective potentials in Fig.~\ref{fig:potentials}. In  the large curvature limit, one can notice that $V'(\phi) \simeq R/3$, and so the equilibrium condition in (\ref{equilibrium}) implies $R \simeq -8\pi G\, T$ just as it is in GR.
As a consistency check, in Fig.~\ref{fig:equilHS} we tested the validity of the general relativistic limit in the case of a non-trivial value of the density contrast in the Hu-Sawicki model for a configuration that remains fully screened. This result remains valid for different values of the density contrast and further extends to the Starobinsky model. 
 
So far, we discussed a few cases where we notice a smooth evolution into a non-trivial static solution. Nonetheless, the existence of static hair solutions is not enough to ensure smooth convergence to them. In Fig.~\ref{fig:potentials}, we find the field values where infinite curvature is reached are close to the minima of the potential for different choices of model parameters, which is consistent with the results in Refs.~\cite{Frolov:2008uf, Appleby:2009uf}. Small field excursions from the minimum are sufficient to form curvature singularities outside the event horizon, and it can happen dynamically.

\begin{figure}
\begin{center}
\includegraphics[width=0.42\textwidth]{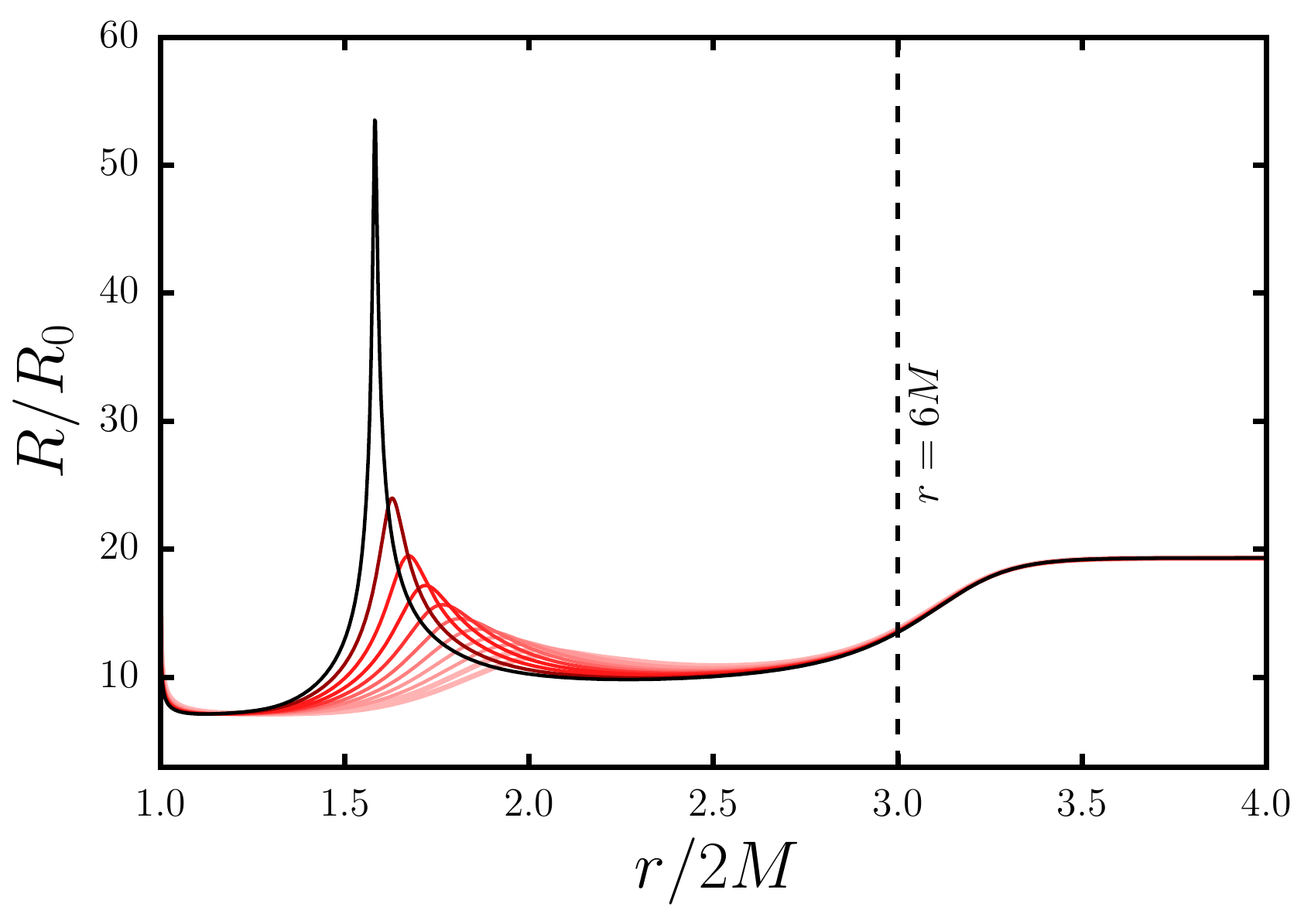}
\caption{Formation of a naked singularity for $n=3$ and $\alpha=4.31$ in the Hu-Sawicki model.}
\label{fig:nakedHS}
\end{center}

\begin{center}
\includegraphics[width=0.42\textwidth]{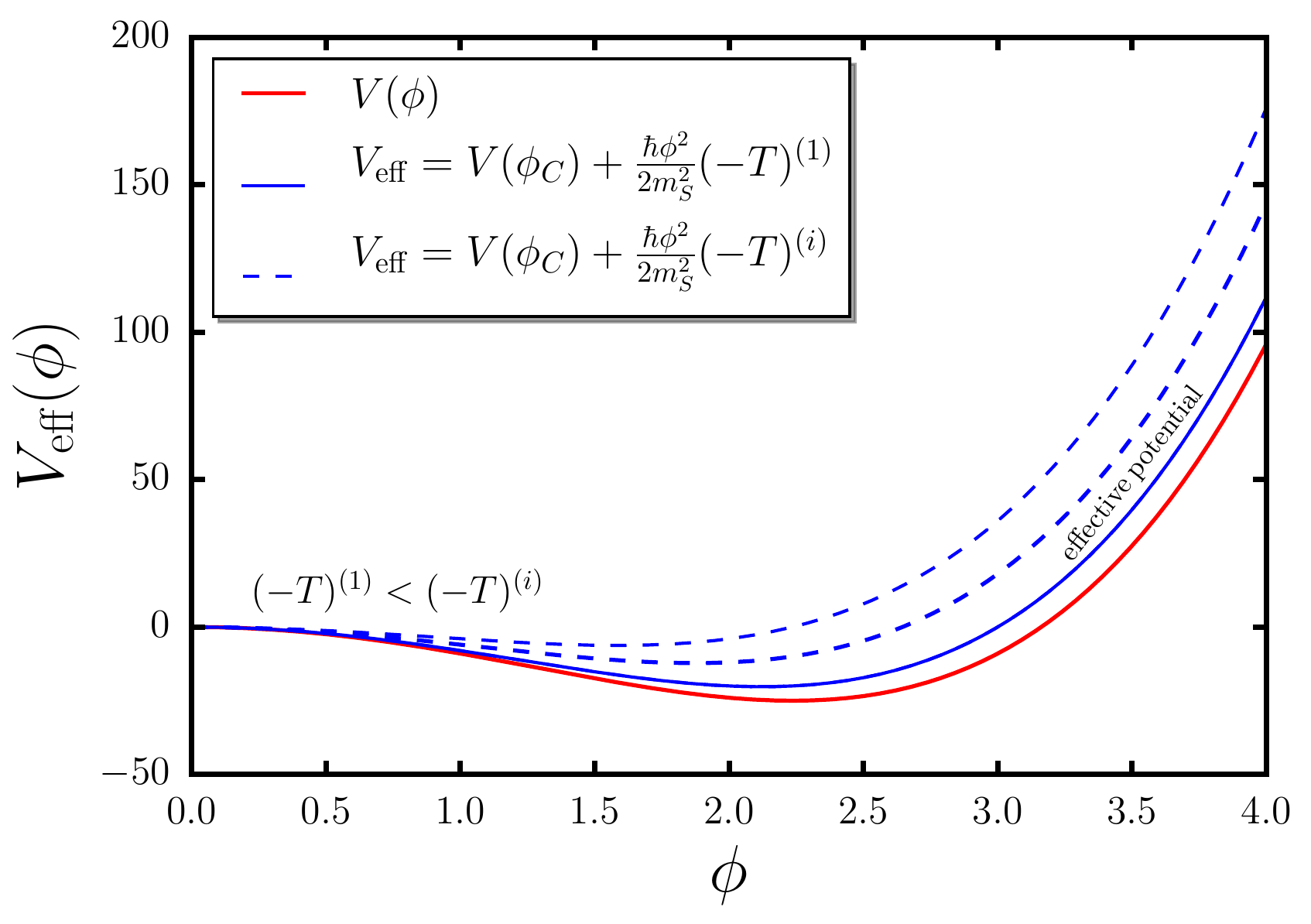}
\caption{Effective potential of the symmetron model. $\phi_*$ goes to zero as the matter density grows.}
\label{fig:pot_symm}
\end{center}
\end{figure}

Equilibrium solutions for the field are defined by locating the ``valleys'' of $V_{\text{eff}}(\phi)$, which in these cases are not distant from reaching the point of infinite scalar curvature. Further modifications of the functional form of $f(R)$, for example the addition of a term proportional to $R^2$, can create a potential wall which will shield the field from reaching singularity. We will explore this alternative in the next section. 

In Fig.~\ref{fig:nakedST}, we show the dynamical formation of a curvature singularity outside the black hole's event horizon. We chose $n=3$ and $\lambda=2.1$ as parameters for the Starobinsky model. Similarly, in Fig.~\ref{fig:nakedHS}, we pick $n=3$ and $\alpha=4.31$ to form a naked singularity in the Hu-Sawicki model. The profiles of field $\phi$ themselves do not diverge or show irregular behaviour while the singularities are formed, but the curvature $R$ goes singular under algebraic inversion $R \equiv R(\phi)$ at $\phi=0$. The singularity is rather weak, and most likely of integrable type, but it is nonetheless a naked singularity formed in evolution of the regular data, which shows cosmic censorship conjecture is violated in these models.

\section{Resolving curvature singularities in $f(R)$ theories}
\label{sec:singularities}
\begin{figure*}
\centering
\subfigure{
\includegraphics[width=0.423\textwidth]{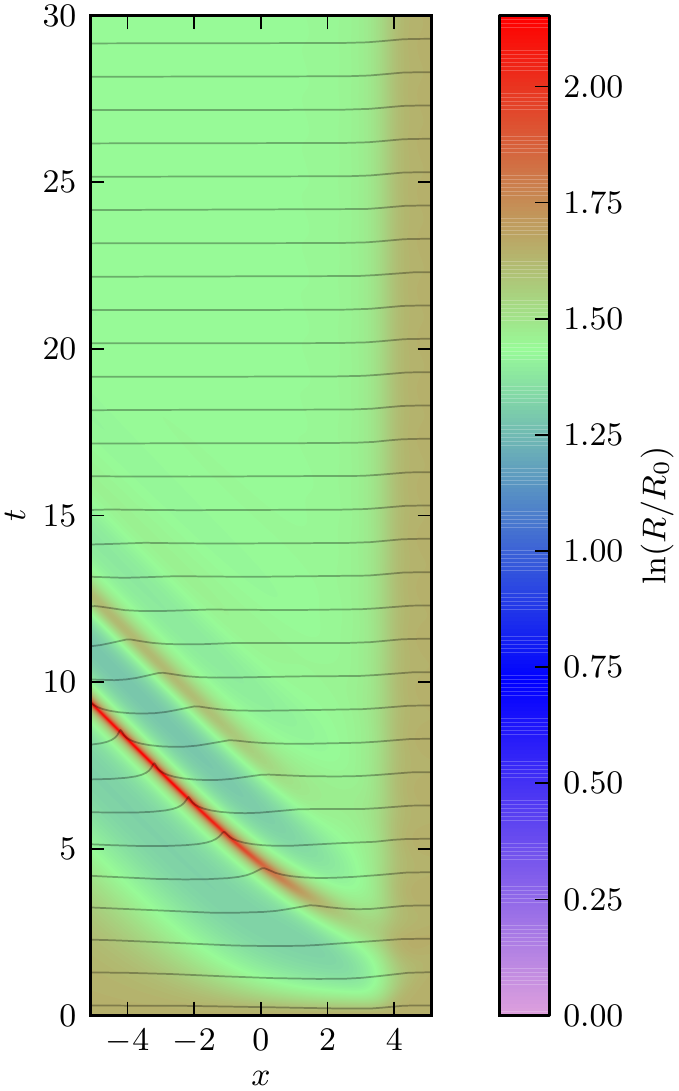}} \,
\subfigure{
\includegraphics[width=0.4125\textwidth]{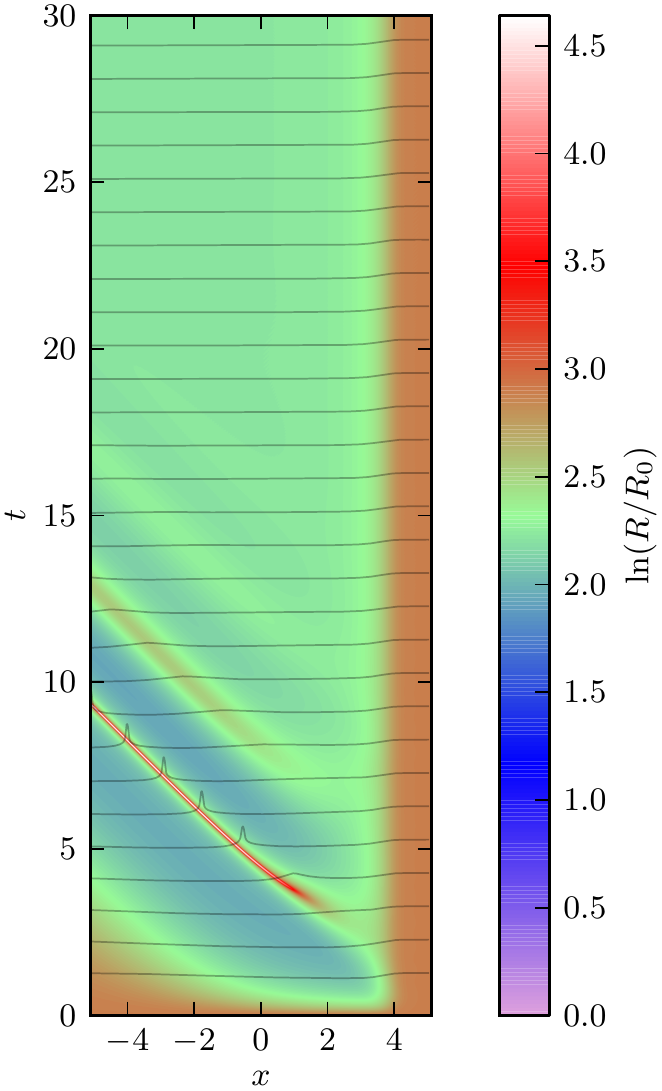}}
\caption{\label{fig:peaks} Dynamical resolution of the curvature singularities found in Figs.~\ref{fig:nakedST} and \ref{fig:nakedHS} after the definitions in \eqref{eqn:fnewstar} and \eqref{eqn:fnewhusawicki}. Left panel: evolution of the curvature cusp for $n=3$ and $\lambda=2.1$ in the Starobinsky model. Right panel: evolution of curvature peak for $n=3$ and $\alpha=4.31$ in the Hu-Sawicki model. In both cases, the formation and absorption of the (approximately null) curvature peaks does not interfere with the stable evolution of the curvature profile.}
\end{figure*}

In Figs.~\ref{fig:nakedST} and \ref{fig:nakedHS}, we observed the formation of integrable curvature peaks in the Starobinsky and Hu-Sawicki models. In both cases, these were located in $r\in(r_g;r_{\mathrm{ISCO}}]$ and appeared as a consequence of small field excursions from the potential minima reaching the field value corresponding to infinite curvature. In this section, we briefly discuss that the addition of an extra ``mass'' term in the functional form of $f(R)$
\begin{eqnarray}
\label{eqn:fnewstar}
&\tilde{f}_{\text{S}}=\displaystyle{R+\lambda\left[\frac{1}{\left(1+(R/R_0)^2\right)^n}-1\right]R_0}+\frac{\mu}{R_0}R^2,\\
\label{eqn:fnewhusawicki}
&\tilde{f}_{\text{HS}}=\displaystyle{R-\frac{\alpha(R/R_0)^nR_0}{1+\beta(R/R_0)^n}}+\frac{\mu}{R_0}R^2,
\end{eqnarray}
is enough to remediate the divergencies appearing in both models for the same choices of model parameters chosen in the previous section, as suggested in Refs.~\cite{Kobayashi:2008wc, Appleby:2009uf}. Here the singular point is avoided by adding an infinite barrier that regularizes the potential and its derivatives. In the same reference, it is possible to find constraints of the value of $\mu$ mostly related with the expected decay time of the scalarons in cosmological scales. Here we chose $\mu = 10^{-6}$ to leave the low-curvature features of the model unaffected, while putting a cap at moderate curvature values to avoid numerical dynamical range issues which might cloud the discussion. Note that the extra mass term does not appear in the definition of $V'(\phi)$ described in \eqref{eqn:fReom}, regardless of any choice of $\mu$. In Fig.~\ref{fig:peaks}, we evaluated the dynamics of the scalar curvature in tortoise coordinates considering $n=3$ and $\lambda=2.1$ as parameters for the Starobinsky model and $n=3$, $\alpha=4.31$ for the Hu-Sawicki model. These choices produced unstable evolution truncated in the large curvature regime when using $f_{\mathrm{S}}$ and $f_{\mathrm{HS}}$. After adding the corrections in \eqref{eqn:fnewstar} and \eqref{eqn:fnewhusawicki}, the curvature peaks are limited and absorbed without reaching infinite values, with evolution towards an equilibrium configuration continuing without further inconveniences. However, even when the evolution of the curvature peaks is more benign these are still formed outside the event horizon. Thus, if such cusps are not observed, their existence in the model will pose constraints not only for specific choices of initial conditions or parameters, but for the entire subspace of the model parameters connected to the troublesome region by the renormalization flow. For the Hu-Sawicki model, the parameter scaling \eqref{eqn:invariance} that leaves action invariant will involve $\mu$, nonetheless it is still possible to find a family of parameters with similar curvature features starting from only one solution.     
   
\section{Scalar accretion for the symmetron model}
\label{sec:symmetron}

\subsection{Action and equations of motion}

We now turn to the symmetron model described by a scalar-tensor action of the form
\begin{equation}
\label{eqn:actsymm}
\begin{split}
S = \int \biggl[ \frac{R}{16\pi G} &- \frac{1}{2} (\nabla \phi)^2 - V(\phi)\biggr] \sqrt{-g}\, d^4x\\
& + S_m [A^2(\phi) g_{\mu \nu}, \psi],
\end{split}
\end{equation}
where $g_{\mu\nu}$ is now in the Einstein frame metric and $\psi$ are the matter fields minimally coupled to the Jordan metric $A^2(\phi) g_{\mu\nu}$. For simplicity, we will describe the dynamics of the scalar $\phi$ in the Einstein frame as in Ref.~\cite{Davis:2014tea}. The symmetron is then modelled with a potential
\begin{equation}
\label{eqn:SymmPot}
V(\phi) = V_0 - \frac{\mu}{2} \phi^2 + \frac{\lambda}{4}\phi^4,
\end{equation}
and a coupling function
\begin{equation}
\label{eqn:SymmCoup}
A(\phi) = 1+ \frac{\hbar}{2}\frac{\phi^2}{m_{\text{S}}^2},
\end{equation}
where $\lambda$ and $\mu$ are positive coupling constants and $m_{\text{S}}^2$ is a high mass scale that suppresses any contributions higher or equal than $\mathcal{O}(\phi^4/m_{\text{S}}^4)$. The equation of motion for the Einstein frame metric are
\begin{equation}
\label{eqn:einsttensoreqmovsymm}
R_{\mu\nu}-\frac{1}{2}Rg_{\mu\nu} = 8\pi G \left( T^{[\phi]}_{\mu \nu}+T^{[m]}_{\mu \nu} \right),
\end{equation}
while for the real scalar $\phi$ the equation of motion reads
\begin{equation}
\label{eqn:symmetron}
\Box\phi = V_{\text{eff}}^{\prime}(\phi).
\end{equation}
The effective potential $V_{\text{eff}} \equiv V(\phi) - T A(\phi)$ is defined as
\begin{equation}
V_{\text{eff}}(\phi)=\tilde{V}_0+\frac{1}{2}\left(-\frac{\hbar T}{m_{\text{S}}^2}-\mu\right)\phi^2+\frac{\lambda}{4}\phi^4,\label{eqn:Veffsymmetron}
\end{equation}
and, as in the case of $f(R)$ theories, it is dependent on the environment through the trace of stress-energy tensor $T \equiv T^\mu_\mu$. The coefficient of $\phi^2$ in equation \eqref{eqn:Veffsymmetron} changes sign depending on the magnitude of $T$, and determines the shape of the symmetron effective potential. In case of an environment made solely of dust ($T=-\rho$), we can define $\rho_{\text{crit}} = \mu m_{\text{S}}^2/\hbar$ such that for energy densities $\rho < \rho_{\text{crit}}$ the effective potential becomes a shape of a mexican hat. For vanishing density, the two minima are at $\phi = \pm\phi_*$ with $\phi_* \equiv \sqrt{\mu/\lambda}$. In regions of high density, $\rho > \rho_{\text{crit}}$, there is one single minimum at $\phi = 0$. Hence $A(\phi)= 1$ for high densities and the field decouples from matter. This mechanism allows the symmetron model, with the proper parameters, to pass the solar system tests of GR. In the case of this model, the difference between the field dynamics in high density regions described in the Jordan and in the Einstein frame is not substantial because of the quadratic dependence of the conformal factor on $\phi/m_{\text{S}}$.

\subsection{Accreting symmetrons}
\begin{figure*}
\centering
\subfigure{
\includegraphics[width=.43\textwidth]{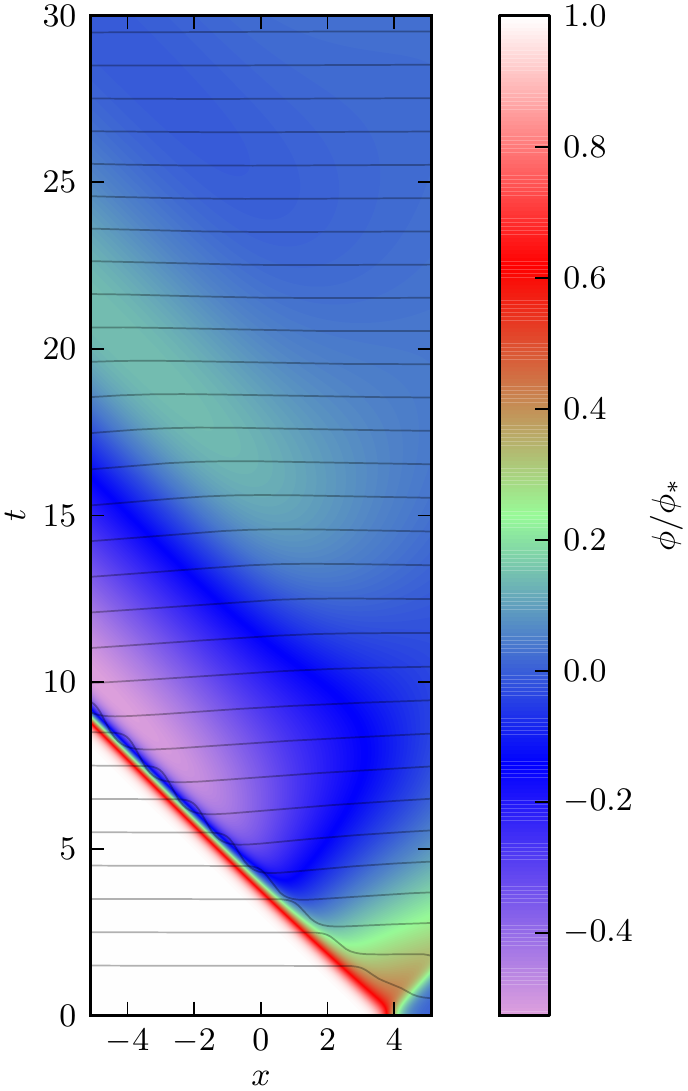}} \,
\subfigure{
\includegraphics[width=.4244\textwidth]{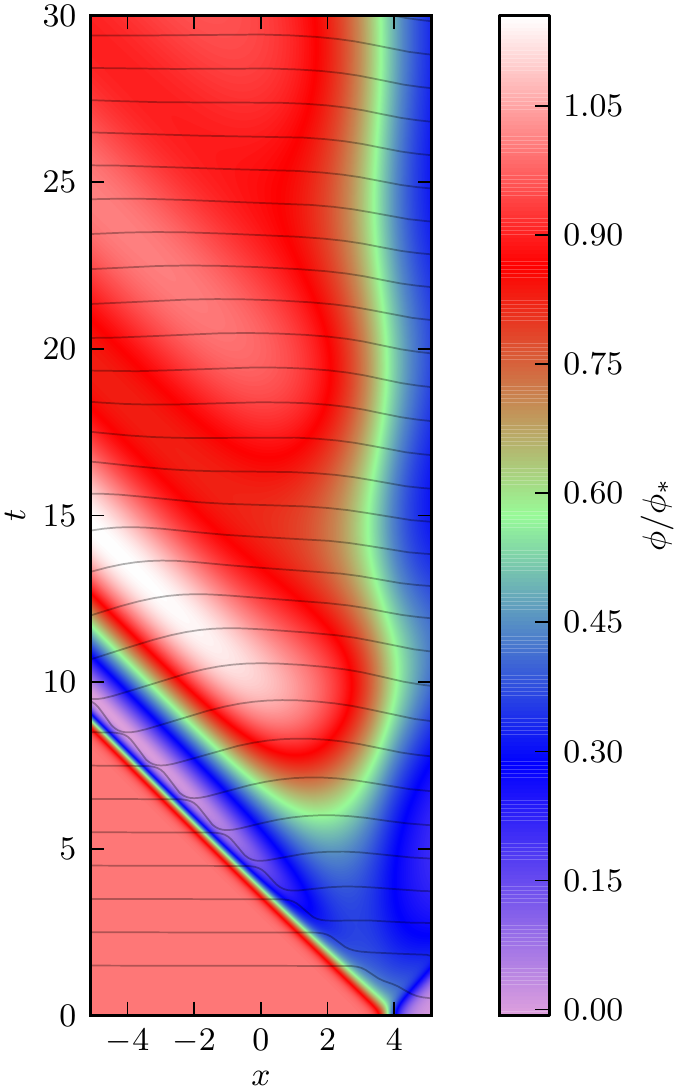}}
\caption{\label{fig:evol_symm} Evolution of the symmetron field for two sets of model parameters in tortoise coordinates. Left panel: Convergence towards a trivial field profile for $\lambda=10^{-2}/\hbar$, $\mu=8\times 10^{-4}/r_g^2$ and $m_{\text{S}}^2=10^{-3} \hbar^2/r_g^2$. Right panel: Non-trivial hair solution for $\lambda=10^{2}/\hbar$, $\mu=8\times 10^{-1}/r_g^2$ and $m_{\text{S}}^2=\hbar^2/r_g^2$.} 
\end{figure*}

\begin{figure}
\begin{center}
\includegraphics[width=0.46\textwidth]{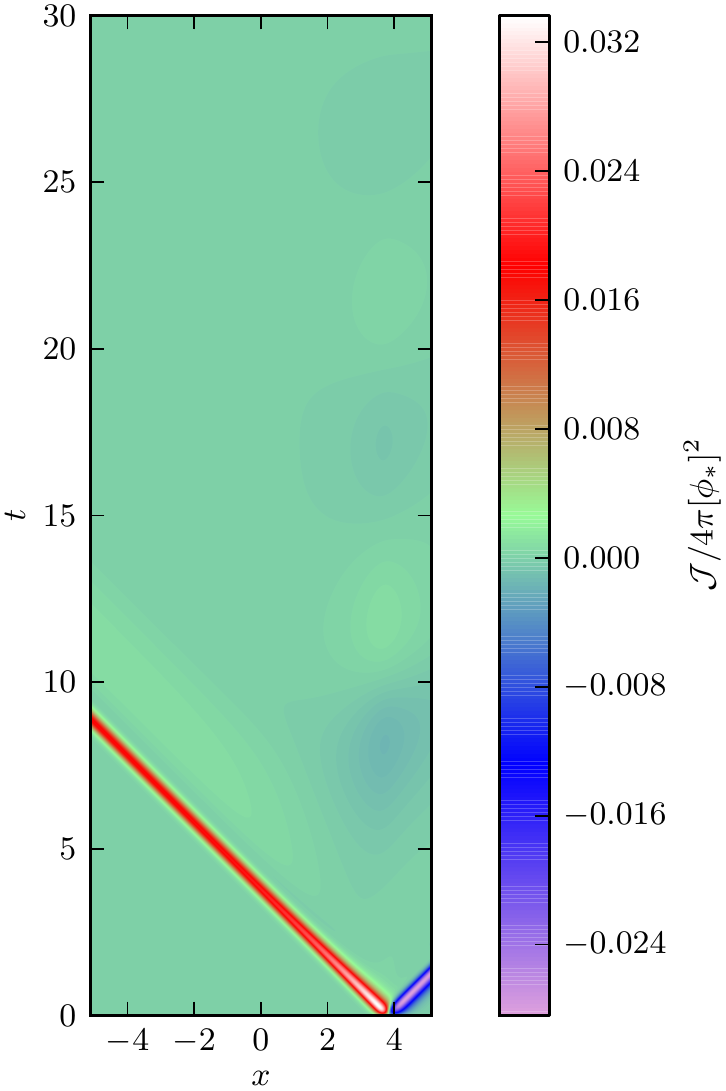}
\caption{Scalar field flux in the symmetron field evolution. Initial transient splits into ingoing and outgoing waves, which are transported to horizon and spatial infinity without attenuation. Overall flux settles to zero soon after initial transient.}
\label{fig:flux_symm}
\end{center}
\end{figure}

In this section we evaluate the dynamics of the accreting symmetrons. Previously, we discussed a simplified model equipped with spontaneously broken $\mathbb{Z}_2$ symmetry and an environmentally dependent mass. The shape of the potential and the effects of the coupling with matter are shown in Fig.~\ref{fig:pot_symm}. In regions of high matter overdensities, the only stable field configuration is $\phi=0$; however, as we discussed in the previous section, the field takes a non-zero vacuum expectation value in zones of lower energy density. In consequence, considering the mass distribution in \eqref{eqn:rho}, the initial radial profile used is
\begin{equation}
\phi(r,t=t_0)= 
     \begin{cases}
       \phi_*, &\quad r<r_{\text{ISCO}}\\
       0, &\quad r\ge r_{\text{ISCO}},
     \end{cases}\label{eqn:symmfield}
\end{equation}
which can be smoothed by using an hyperbolic tangent in the same manner as in the matter profile. An equivalent but more physically motivated way to choose the initial field configuration is to place it in equilibrium $V_{\text{eff}}'(\phi)=0$ for a given matter distribution. Once again we will assume $\dot{\phi}(r,t=t_0)=0$ for initial field velocity. With these initial conditions and choosing $\sigma=0$, we compute the evolution of the field profile for two different sets of model parameters: (a) $\lambda=10^{-2}/\hbar$, $\mu=8\times 10^{-4}/r_g^2$, $m_{\text{S}}^2 = 10^{-3} \hbar^2/r_g^2$ and (b) $\lambda=10^{2}/\hbar$, $\mu = 8\times 10^{-1}/r_g^2$, $m_{\text{S}}^2 = \hbar^2/r_g^2$. In Fig.~\ref{fig:evol_symm}, we calculated the evolution of the field towards equilibrium for both choices. Depending on parameters chosen, symmetrons do not always form non-trivial static hair solutions. The flux of the symmetron field ${\cal J} \equiv 4\pi r^2\,\phi_{, x}\phi_{,t}$ is shown in Fig.~\ref{fig:flux_symm}. All the ingoing scalar fluxes calculated here settle to zero smoothly after initial transient.
 

\begin{figure}
\begin{center}
\includegraphics[width=0.42\textwidth]{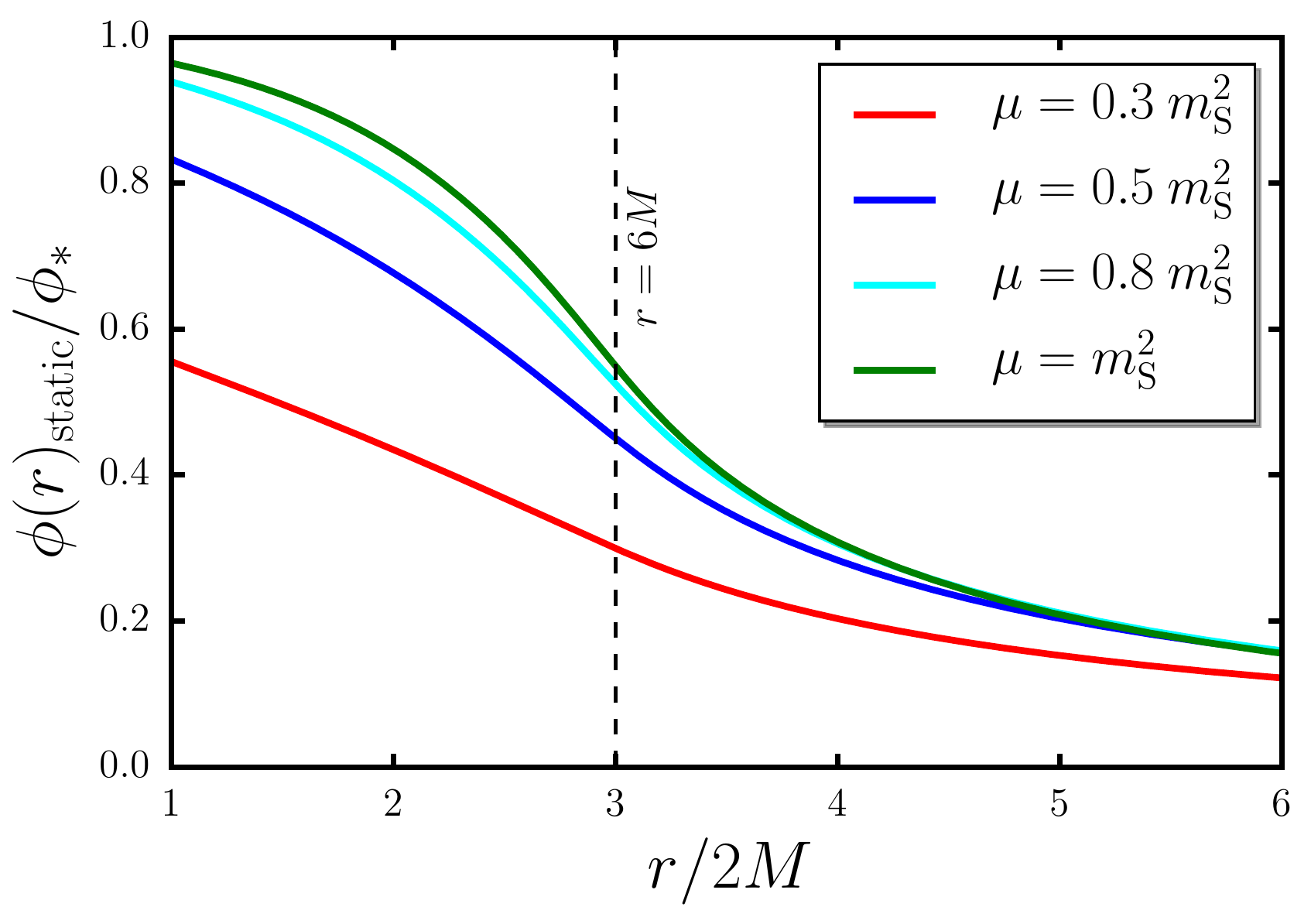}
\caption{Static solutions of the symmetron field for different values of $\mu$, we set $\sigma=0$ as a constant.}
\label{fig:symmmus}
\end{center}
\end{figure}

\begin{figure}
\begin{center}
\includegraphics[width=0.42\textwidth]{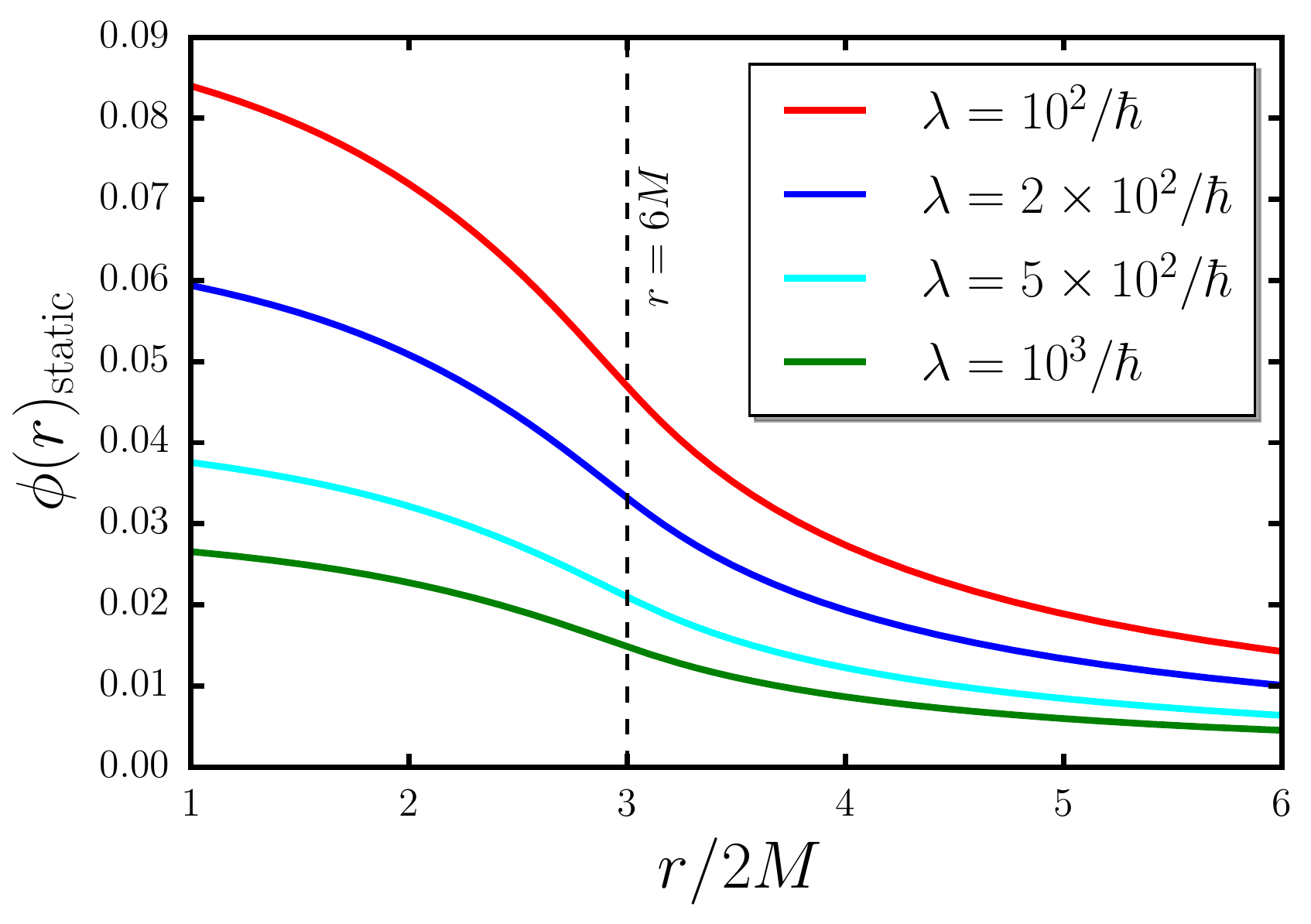}
\caption{Static solutions of the symmetron field for different values of $\lambda$, $\mu=0.8m_\text{S}^2$ was held as a constant. $\lambda$ acts as an overall normalization constant for the solution.}
\label{fig:symlambda}
\end{center}
\end{figure}

\begin{figure}
\begin{center}
\includegraphics[width=0.42\textwidth]{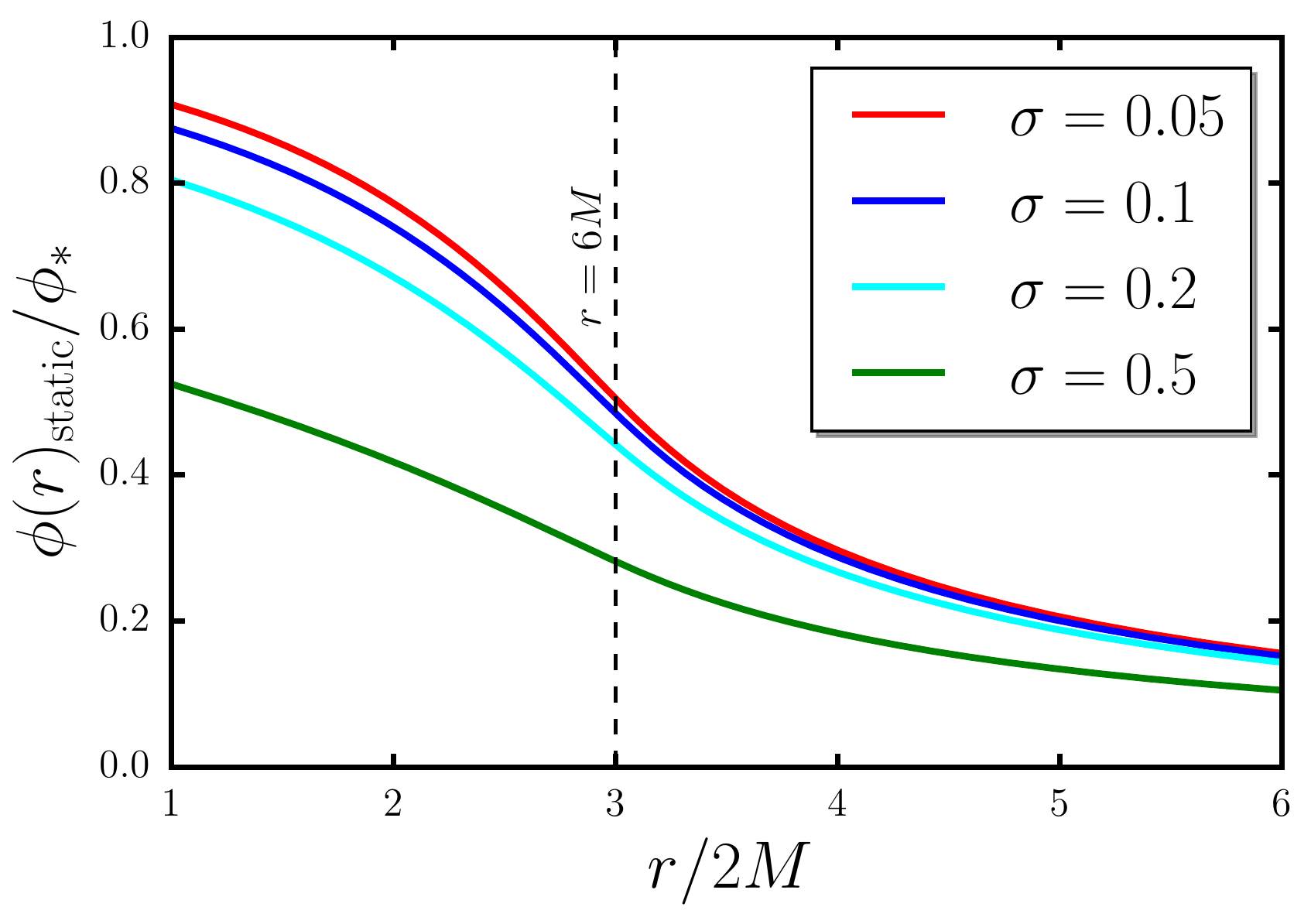}
\caption{Field profiles for different values of $\sigma$ in the symmetron model, $\mu=0.8m_\text{S}^2$ and $\lambda=10^2$ were held as constants.}
\label{fig:symmsigma}
\end{center}
\end{figure}

The difference in evolution lies in the contribution of the matter source to the effective potential: In the left panel of Fig.~\ref{fig:flux_symm}, we supressed the effects of the ``external force'' driven by the static matter density. Therefore this limit case is consistent with the standard no-hair theorem. However, that is not the case for the model depicted in the right panel, where the mass of the black hole is the same as the coupling parameter $m_{\text{S}}$. In accordance with our description in the previous subsection, we will explore the cases in which we can find non-trivial static solutions. From Fig.~\ref{fig:pot_symm}, we can recognize at least one of equilibrium field configurations which correspond to the different vacua for a given shape of the deformed ``Mexican hat'' potential. In Figs.~\ref{fig:symmmus}-\ref{fig:symmsigma}, we compute the static solutions for different values of the model parameters.

\section{Accretion of the Ratra-Peebles chameleons}
\label{sec:ratrapeebles}

\subsection{Equations of motion and setup}
We now study the prototypical example of the chameleon screening, defined in the Einstein frame by the same scalar-tensor action as in \eqref{eqn:actsymm}, with the Ratra-Peebles potential
\begin{equation}
V(\phi) = V_0+\frac{\gamma^{n+4}}{\phi^n}
\label{eqn:ratrapeebles}
\end{equation}
and an approximate coupling function
\begin{equation}
A(\phi) \approx 1+\frac{\varepsilon\phi}{m_{\text{C}}},
\label{eqn:rpcoupling}
\end{equation}
where $\gamma$ and $\varepsilon$ are positive model parameters and $m_{\text{C}}$ plays the role of a high mass scale where the screening is effective. Here we also choose to work in the Einstein frame being consistent with the procedures followed in Ref.~\cite{Davis:2014tea}. The equations of motion for both the metric and the scalar field are the same as in \eqref{eqn:einsttensoreqmovsymm} and \eqref{eqn:symmetron}, but now the effective potential is given by
\begin{equation}
V_{\text{eff}}(\phi) = V_0+\frac{\gamma^{n+4}}{\phi^n}+\frac{\varepsilon\phi}{m_{\text{C}}}(-T).
\label{eqn:Veffchameleon}
\end{equation}
This is another case of an effective potential dependent on the environment. Due to the term proportional to $T$, it is possible to find an equilibrium configuration for the field from $V'_{\text{eff}}(\phi)=0$
\begin{equation}
\bar{\phi}_{\text{C}} = \left(n\gamma^{n+4}\,\frac{m_{\text{C}}}{-\varepsilon T}\right)^{\frac{1}{n+1}},
\label{eqn:chamfieldeq}
\end{equation}
in the case of pressureless matter $(T=-\rho)$, we see that the field is suppressed in dense regions (with energies larger than $m_{\text{C}}$) and unscreened in regions with lower densities. The effective mass of the field is given by
\begin{equation}
V''_{\text{eff}}(\bar{\phi}_{\text{C}}) = (n+1)\, n^{-\frac{1}{n+1}}\,\gamma^{-\frac{n+4}{n+1}} \left(\frac{-\varepsilon T}{m_{\text{C}}}\right)^{\frac{n+2}{n+1}},
\label{eqn:effmasscham}
\end{equation}
which becomes larger in dense environments. It is widely known that chameleon screening occurs in a different way as in the case of the symmetron as a consequence of the reduction of the field correlation length due to a larger effective mass of the field. In a low density environment, the minimum can be found at $\phi = \phi_*$ with $\phi_* \equiv \left(n\gamma^{n+4}\,m_{\text{C}}/(\varepsilon\sigma \rho_0)\right)^{\frac{1}{n+1}}$, where $\sigma\rho_0$ is the density at $r<r_{\text{ISCO}}$.

\subsection{Accreting chameleons}

\begin{figure*}
\centering
\subfigure{
\includegraphics[width=.41\textwidth]{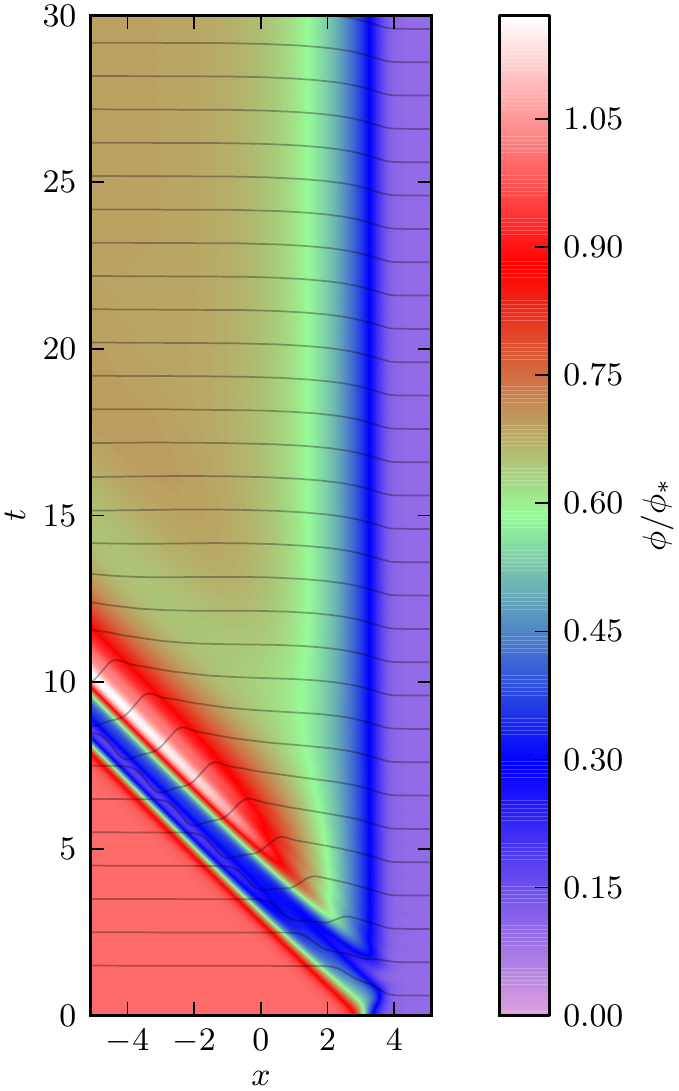}} \,
\subfigure{
\includegraphics[scale = 1.07]{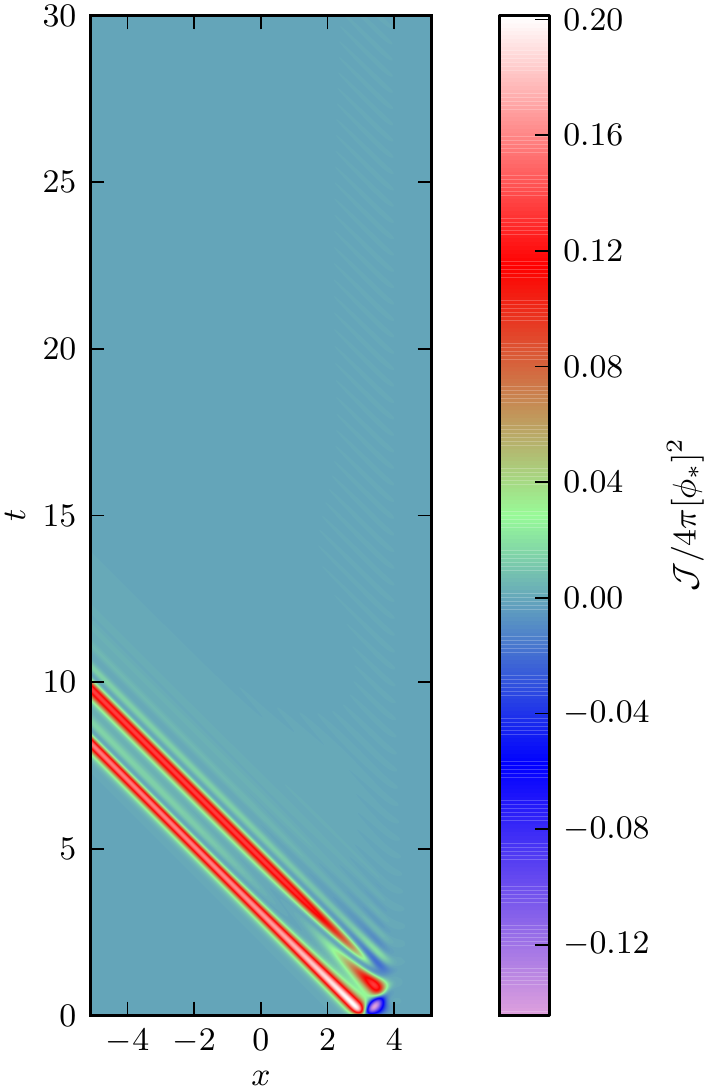}}
\caption{\label{fig:evol_rp} Left panel: Evolution of the Ratra-Peebles chameleon for $n=1$. The field smoothly evolves into a static hair solution after initial transient. Right panel: Scalar field flux of the same evolution. The scalar flux represents how the field gets absorbed into the black hole's event horizon.} 
\end{figure*}

\begin{figure}
\begin{center}
\includegraphics[width=0.41\textwidth]{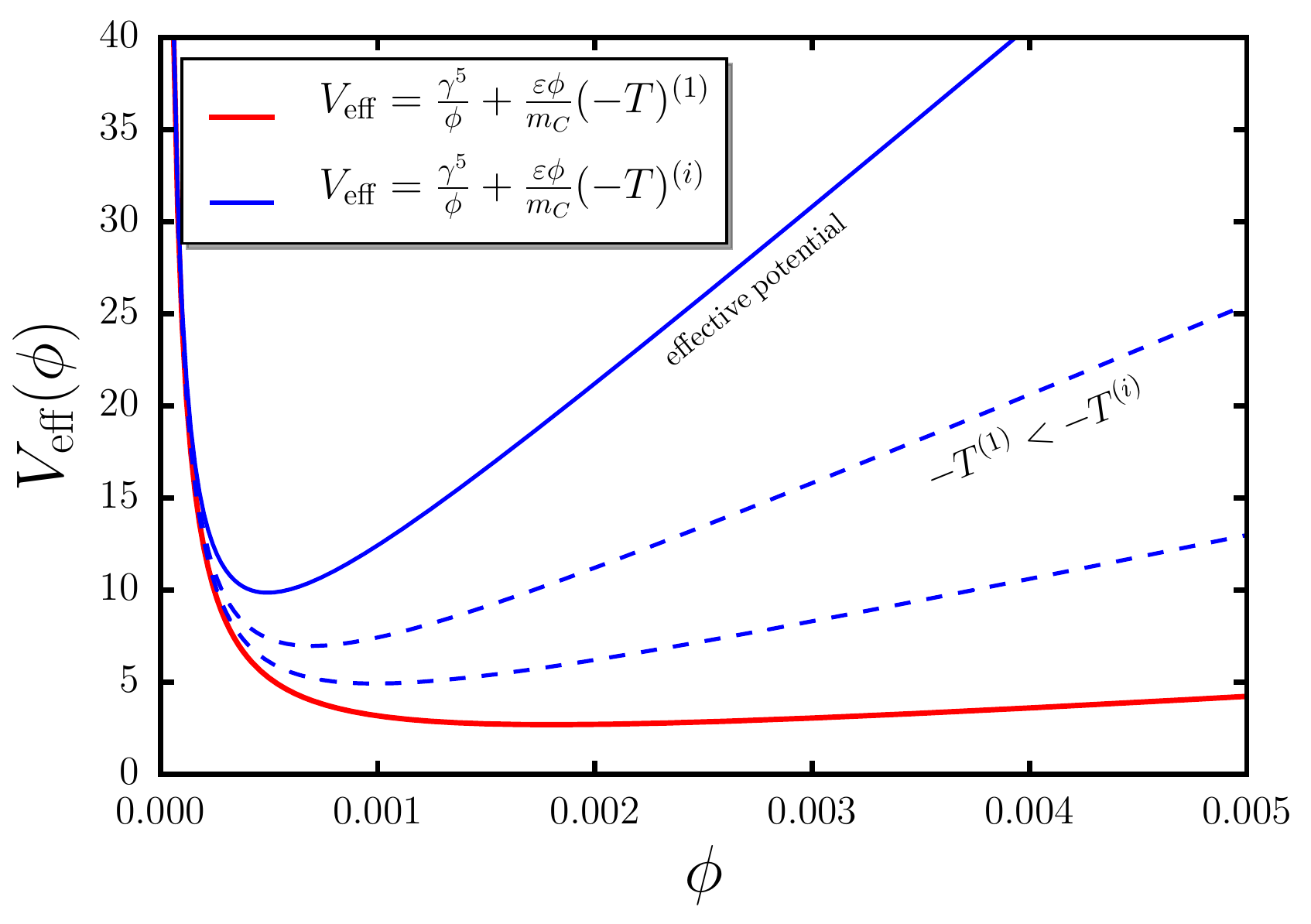}
\caption{Effective potential for the Ratra-Peebles chameleon considering arbitrary spatially constant matter densities and positive model parameters for $n=1$.}
\label{fig:pot_cham}
\end{center}
\end{figure}

Here we evaluate the accretion of the Ratra-Peebles chameleons in the background described in section \ref{sec:matterset}. In Fig.~\ref{fig:pot_cham} we observe the shape of the potential found as a function of the field $\phi$, and the effects of the coupling with matter. The effective mass -- i.e. the concavity of the effective potential -- increases with the environmental matter density of dust.

\begin{figure}
\begin{center}
\includegraphics[width=0.41\textwidth]{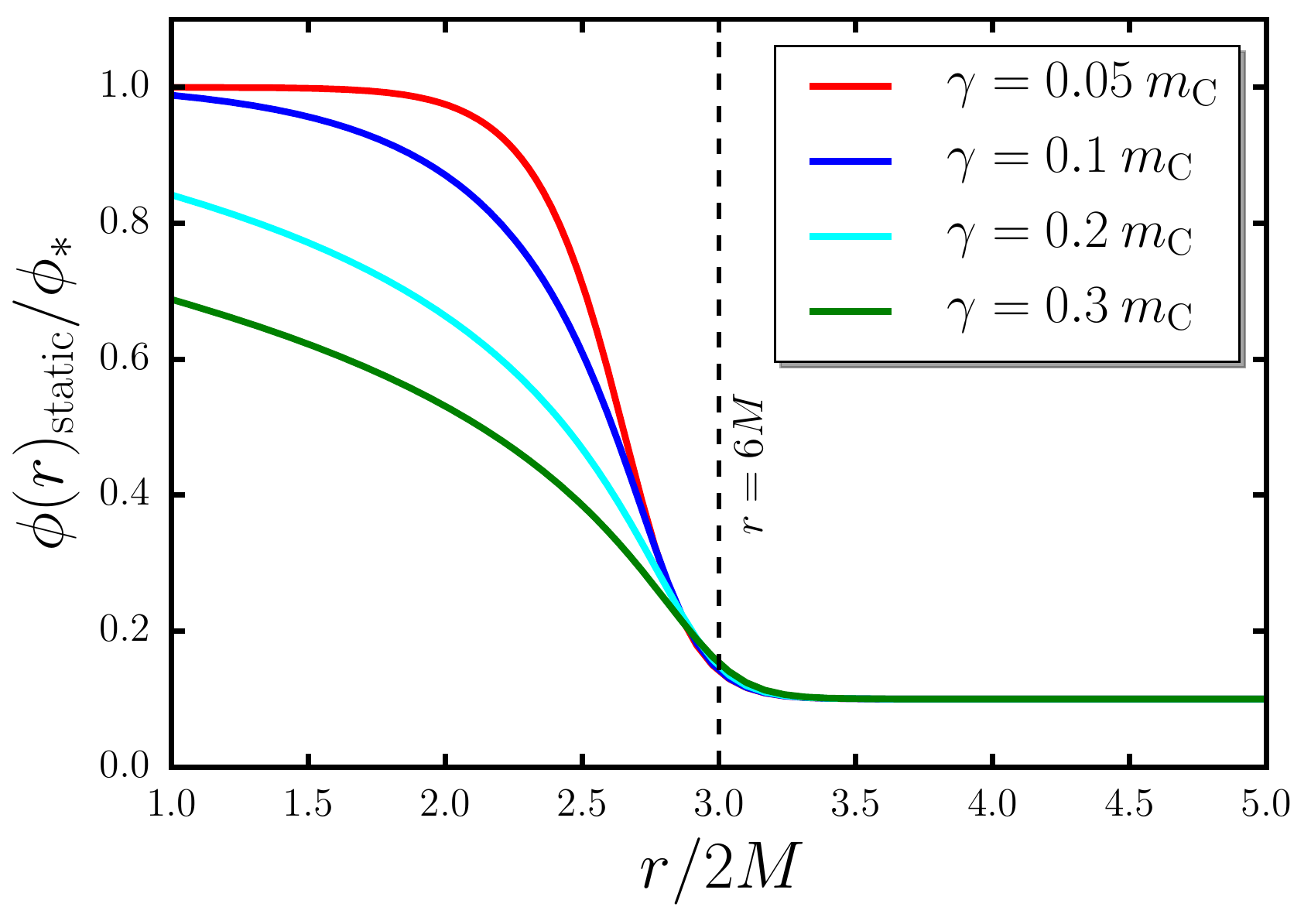}
\caption{Static solutions of the chameleon field for different values of $\gamma$, we set $n=1$ and $\sigma=0.01$ as constants.}
\label{fig:rpgammas}
\end{center}
\end{figure}

\begin{figure}
\begin{center}
\includegraphics[width=0.41\textwidth]{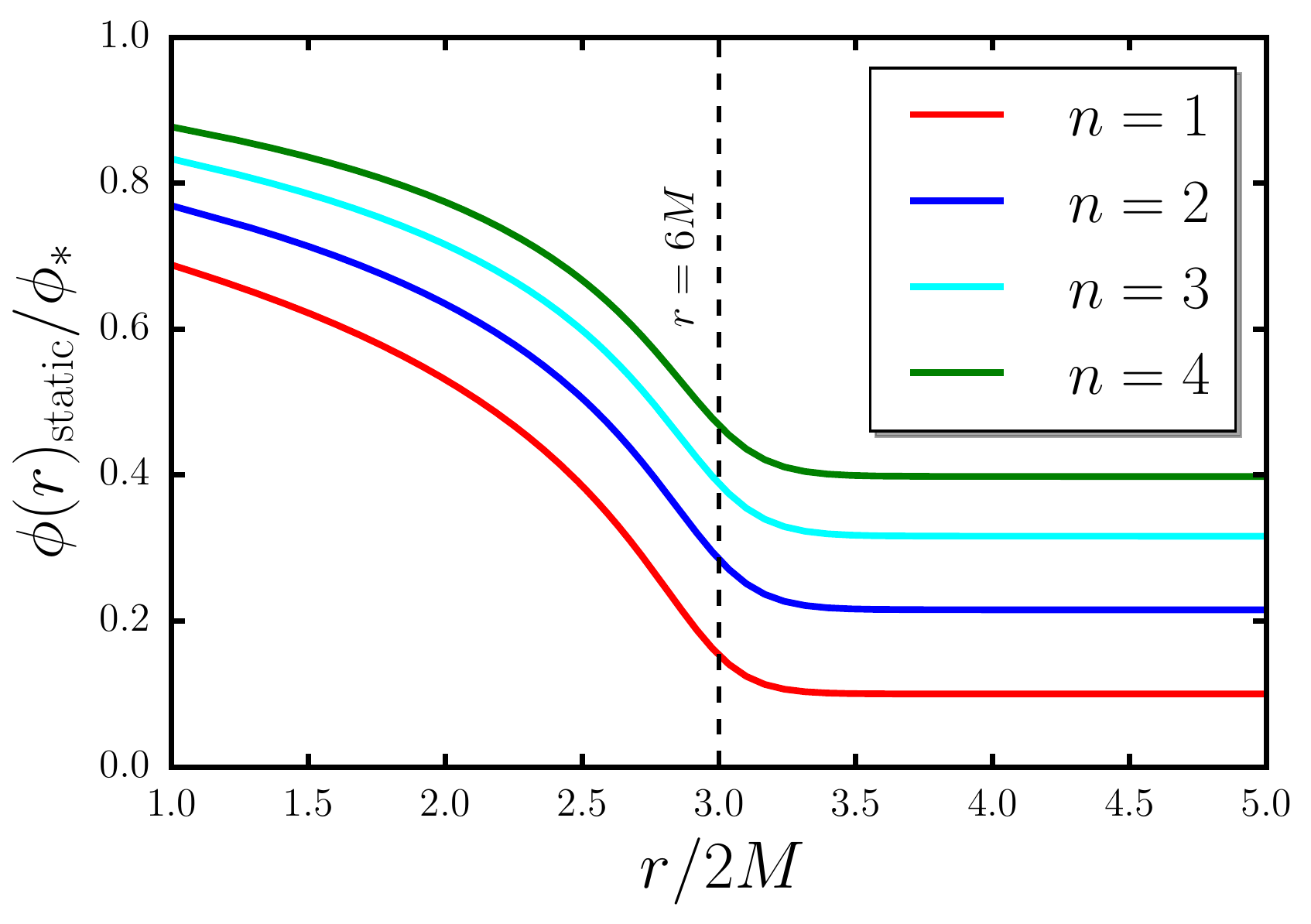}
\caption{Ratra-Peebles chameleons for different values of $n$. This parameter also shifts the initial equilibrium configuration for $\phi$.}
\label{fig:rpns}
\end{center}
\end{figure}

\begin{figure}
\begin{center}
\includegraphics[width=0.41\textwidth]{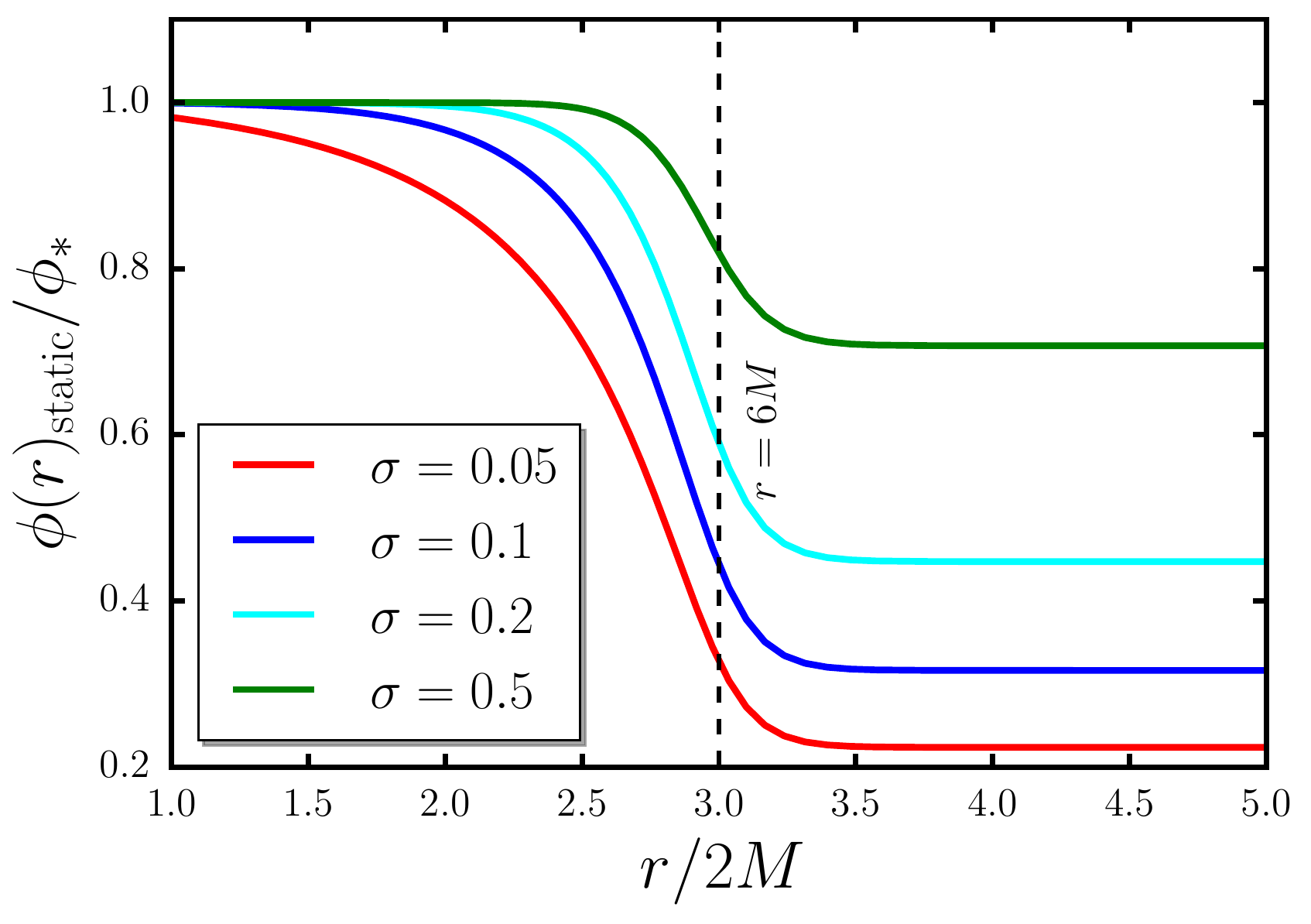}
\caption{Field profiles for different values of $\sigma$ in the chameleon model. Notice the screening at $r<r_{\text{ISCO}}$ of the hair solution for larger values of $\sigma$.}
\label{fig:rpsigmas}
\end{center}
\end{figure}

We derive initial conditions compatible with \eqref{equilibrium} by replacing the matter distribution proposed in \eqref{eqn:rho} into \eqref{eqn:chamfieldeq}, avoiding the value of $\sigma=0$ to not have any divergencies in the initial field profile at $r<r_{\text{ISCO}}$. The constitution of the Ratra-Peebles model offers the possibility of finding the corresponding initial field configuration as a function of the surrounding matter density, the reverse process can also be coded without inconveniences. Assuming static initial conditions and $\sigma=0.01$, $\varepsilon=10^2$, $\gamma=0.3 m_{\text{C}}$ and a surrounding dust density of $-T=\rho_0=3\times10^{-2}m_{\text{C}}/r_g^3$, we find the evolution of the field from initial conditions towards a static solution in the left panel of Fig.~\ref{fig:evol_rp}. From this figure, we can notice that the evolution of the field is not significantly different from our results for the Starobinsky and the Hu-Sawicki models in Fig.~\ref{fig:evolution}, which is consistent with the effects of the chameleon screening in these models. Additionally, the ingoing field flux is depicted in the right panel of the same figure. Here the flux is regular and converges to the limit where there is no other source apart from the static matter distribution. Ingoing flux lines are represented around $x\left(r_{\text{ISCO}}\right)\approx 3.89$. We also present static solutions for different choices of the model parameters. In Figs.~\ref{fig:rpgammas} - \ref{fig:rpsigmas}, we show the changes in shape of the hair profiles for different parameter choices. In this case, the tuning of the parameter in the runaway potential is sensitive to changes in the orders of magnitude: it is simple to suppress the whole contribution this part of the potential by accident, due to the $(n+4)$ power of $\gamma$.



\section{Discussion}\label{sec:conclusions}
In this paper, we describe the dynamics of scalar accretion onto a Schwarzschild black hole in the presence of a static matter distribution modelled by \eqref{eqn:rho}. In particular, we studied the accretion of the extra scalar degrees of freedom appearing in two models of $f(R)$ gravity, the Starobinsky and Hu-Sawicki model, as well as in the symmetron model. Stable convergence to static scalar hair profiles results from varying parameters for each specific model. In the cases of the Starobinsky and Hu-Sawicki model, we can obtain dynamical chameleon solutions with singular curvature outside horizon without requiring an infinite energy budget. In the case of accreting symmetrons, it is not always possible to form a non-trivial static solution since it depends on the strength of the coupling with matter. More concretely, it depends on how large is the energy scale $m_{\text{S}}$ compared to the mass of the black hole. Our results for the field fluxes are included for all the cases we studied, along with the static solutions for different model parameters.

Even when the simulations of astrophysical rotating black holes suggest a large density contrast, we noticed that the formation of non-trivial static solutions does not require an absolute vacuum environment close to the black hole's event horizon, where $r\in(r_g, r_{\text{ISCO}}]$. We did not consider the effects of backreaction of the field in the spacetime solutions since these are small even during the formation of integrable naked singularities.  

We acknowledge the progress made in Refs.~\cite{Davis:2014tea, Davis:2016avf}, wherein approximate analytic expressions for scalar hair solutions in the case of rotating black holes are found. Additionally, this paper discusses the possibility of a non-negligible ratio between the radiated power from extra scalar sources and the quadrupole gravitational radiation in GR, which might be testable by the future generation of gravitational-wave detectors. In our approach, apart from calculating static solutions in different circumstances, we evaluated the scalar accretion dynamically in such a way that it is possible to converge to a hairy or a ``bald'' solution, depending on the model and its parameters. These results also motivate further explorations on the effect of fifth forces confined by screening, surrounded by a non-trivial matter profile for merging binary systems.

The existence of non-trivial field profiles has also been studied for scalars with non-canonical kinetic terms, such as in the Galileons studied in Refs.~\cite{Babichev:2013cya, Babichev:2016fbg}. Those solutions are described in vacuum environments and do not accrete into the black hole's horizon. The form of the equations of motion of Galileon is not semi-linear, and is harder to study numerically, as Galileons propagate with speeds that vary at different locations. Numerical implementation of a dynamical code designed to compute evolution of such scalar fields will be covered in a future project. We are also considering to extend the techniques developed here to full 3D scalar field scattering by a black hole.

\appendix
\section{Numerical Implementation}\label{app:numerical}

\subsection{Scalar field equations of motion}\label{app:numerical:eom}

Equations of motion describing evolution of a scalar field $\phi$ with a (non-linear) self-interaction potential $V(\phi)$ and an external force term ${\cal F}$ propagating on a fixed background spacetime are described by a (semi-linear) PDE
\begin{equation}
  \Box\phi = V'(\phi) - {\cal F},
\end{equation}
where $\Box$ denotes a covariant d'Alembert operator. For a spherically symmetric black hole described by the Schwarzschild metric
\begin{equation}
  ds^2 = -g(r)\, dt^2 + \frac{dr^2}{g(r)} + r^2\, d\Omega^2,
\end{equation}
where $d\Omega^2$ is the metric on a unit sphere, and the metric function $g(r)$ is
\begin{equation}
  g(r) = 1 - \frac{2M}{r},
\end{equation}
the left hand side of the equation of motion is simply
\begin{eqnarray}
  \Box\phi &=& 
  \frac{1}{\sqrt{-g}}\, \partial_\mu \Big( \sqrt{-g}\, g^{\mu\nu}\, \partial_\nu \phi \Big)\nonumber\\ &=&
  - \frac{1}{g(r)}\, \partial_t^2 \phi + \frac{1}{r^2}\, \partial_r \Big( r^2 g(r)\, \partial_r \phi \Big).
\end{eqnarray}
This can be reduced to a one-dimensional wave equation with constant propagation speed by introducing the tortoise coordinate $x$ by $\partial_x = g(r)\, \partial_r$. With this re-definition, the scalar field equation of motion reads
\begin{equation}\label{eq:wave}
  - \partial_t^2 \phi + \frac{1}{r^2}\, \partial_x \Big( r^2\, \partial_x \phi \Big) = g \Big( V'(\phi) - {\cal F} \Big).
\end{equation}
Explicit form of the tortoise coordinate $x$ for the Schwarzschild spacetime can be obtained by integrating
\begin{equation}
  x = \int \frac{dr}{g(r)} = r + 2M\,\ln\left(\frac{r}{2M} - 1\right).
\end{equation}
Tortoise coordinate $x$ is vastly preferable for numerical integration of the wave equation over areal coordinate $r$ since the characteristic speed is constant on the sampled time slice, but the added difficulty with this choice is that accurate $r(x)$ inversion is quite not-trivial numerically, as detailed in Appendix~\ref{sec:tortoise}.

The standard numerical evolution scheme would involve first-order Hamiltonian dynamical system
\begin{equation}
  \dot\phi = \pi, \hspace{1em}
  \dot\pi = \frac{1}{r^2}\, \partial_x \Big( r^2\, \partial_x \phi \Big) - g \Big( V'(\phi) - {\cal F} \Big).
\end{equation}
However, as we will see in the following Section, it is easier to handle absorbing boundary conditions if we rewrite equations of motion in a flux-conservative form by introducing auxiliary variables $u \equiv \partial_t \phi$ and $v \equiv r^2 \partial_x \phi$, so that equations of motion become
\begin{eqnarray}\label{eq:flux}
  - \partial_t u + \frac{1}{r^2}\, \partial_x v &=& g \Big( V'(\phi) - {\cal F} \Big),\nonumber\\
  - \partial_t v + r^2\, \partial_x u &=& 0.
\end{eqnarray}
The first equation is the identical rewrite of the wave equation (\ref{eq:wave}), while the second is the integrability condition requiring that the partial derivatives of $\phi$ commute.

\subsection{Absorbing boundary conditions}\label{app:numerical:pml}

The scalar field degree of freedom $\phi$ asymptotes to a free field evolution near horizon (where $g \rightarrow 0$), and a massive field evolution far away from the black hole (where $V'(\phi) \rightarrow {\cal F}$). Physically, excitations in $\phi$ take infinite amount of time $t$ to reach both boundaries, yet truncating or compactifying the evolution domain for numerical purposes will inevitably lead to spurious reflections unless special care is taken. The best technique to avoid spurious reflections is to introduce absorbing boundary conditions via Perfectly Matched Layers (PMLs) as described in Ref.~\cite{Johnson:wt}, which damp the solution at the boundaries while guaranteeing identically vanishing reflection coefficient at the absorption layer. This is achieved by analytic continuation of the equations of motion into the complex domain
\begin{equation}
  x \rightarrow x + i f(x), \hspace{1em}
  \partial_x \rightarrow \frac{\partial_x}{1 + i f'(x)} \equiv
  \frac{\partial_x}{1 + \frac{\gamma(x)}{\partial_t}},
\end{equation}
which turns the oscillatory travelling waves $e^{ikx-i\omega t}$ into exponentially decaying functions of $x$ instead. To make attenuation length independent of $\omega$, frequency dependent contour deformation $f' = \gamma(x)/\omega$ is chosen and $i/\omega$ is transformed back into explicit integration operator in the time domain. Applying this idea to the scalar field equations of motion in flux-conservative form (\ref{eq:flux}) for an arbitrary damping function $\gamma(x)$, we obtain
\begin{eqnarray}
  - (\partial_t + \gamma) u + \frac{1}{r^2}\, \partial_x v &=& \left(1 + \frac{\gamma(x)}{\partial_t}\right) \Bigg[g \Big( V'(\phi) - {\cal F} \Big)\Bigg],\nonumber\\
  - (\partial_t + \gamma) v + r^2\, \partial_x u &=& 0.
\end{eqnarray}
To turn the inverse time evolution operator $\partial_t^{-1}$ into a differential equation form, introduction of a third auxiliary variable $w$ is in order. With re-definition $u \rightarrow u+w$, the non-reflecting PML equations of motion then become
\begin{subequations}\label{eq:pml}
\begin{eqnarray}
  \partial_t \phi &=& u-w,\\
  \partial_t u &=& \frac{1}{r^2}\, \partial_x v - \gamma u,\\
  \partial_t v &=& r^2\, \partial_x (u-w) - \gamma v,\\
  \partial_t w &=& g \Big( V'(\phi) - {\cal F} \Big).
\end{eqnarray}
\end{subequations}
The damping function $\gamma(x)$ can be quite arbitrary, but it should have compact support near the boundaries to not affect the evolution in the interior, and have sufficient support and magnitude to absorb the impinging waves which hit the boundary during the expected evolution.

\subsection{Spectral basis}\label{app:numerical:basis}

As the scalar field is usually quite stiff and does not form shocks in the course of evolution, the method of choice to evaluate derivative operators is spectral, as described in Ref.~\cite{Boyd}. Compactifying the tortoise coordinate $x$ on a scale $\ell$
\begin{equation}
  y = \frac{x}{\sqrt{x^2+\ell^2}} \equiv \cos\theta, \hspace{1em}
  \frac{x}{\ell} = \frac{y}{1-y^2} = \cot\theta
\end{equation}
and introducing a Chebyshev basis on interval $y \in [-1,1]$
\begin{eqnarray}
  T_n &=& \cos(n\theta),\\
  \partial_x T_n &=& \frac{n}{\ell}\, \sin(n\theta)\, \sin^2\theta,\nonumber\\
  \partial_x^2 T_n &=& \frac{n}{\ell^2} \Big(n \cos(n\theta) + 2 \cot\theta \sin(n\theta)\Big) \sin^4\theta,\nonumber
\end{eqnarray}
we arrive at the spectral representation of the solution
\begin{equation}
  \phi(x) = \sum\limits_n c_n T_n(y)
\end{equation}
truncated to a finite number of modes. While Galerkin method to discretize equations of motion can be employed, the simplest method to evaluate derivative operators is pseudo-spectral, where equations of motion are solved on a Gauss-Lobatto grid
\begin{equation}
  \theta_i = \left(n-i+\frac{1}{2}\right)\,\frac{\pi}{n}, \hspace{1em}
  x_i = \ell\cot\theta_i.
\end{equation}
One does not have to explicitly find coefficients $c_n$ to evaluate the derivative operators of a function $\phi(x)$ sampled on a collocation grid $x_i$. Instead, derivative operators like ${\cal D}_{ij}$ and ${\cal L}_{ij}$ can be found in advance by solving linear matrix equations
\begin{subequations}
\begin{eqnarray}
  \sum\limits_j {\cal D}_{ij} T_n(x_j) &=& \partial_x T_n(x_i),\\
  \sum\limits_j {\cal L}_{ij} T_n(x_j) &=& \left(\partial_{x} + \frac{2g}{r}\right) \partial_x T_n(x_i),
\end{eqnarray}
\end{subequations}
and so on for every basis function $T_n$ evaluated at all nodes $x_i$.

\subsection{Gauss-Legendre integrator}\label{app:numerical:gl}

Packing the scalar field variables $\phi,u,v,w$ evaluated at the collocation grid points $x_i$ into a state vector $\vec{y} \equiv \{ \phi(x_i), u(x_i), v(x_i), w(x_i) \}$, the wave equation (\ref{eq:pml}) reduces to an autonomous dynamical system
\begin{equation}
  \frac{d\vec{y}}{dt} = \vec{f}(\vec{y}),
\end{equation}
which can be integrated by an implicit Runge-Kutta method, as presented in Ref.~\cite{Butcher:1964fx}
\begin{equation}
  \vec{y} \rightarrow \vec{y} + \Delta t \cdot \sum\limits_i b_i \vec{g}^{(i)},
\end{equation}
where the trial directions $\vec{g}^{(i)}$ are defined by
\begin{equation}
  \vec{g}^{(i)} = \vec{f}\left(\vec{y} + \Delta t \cdot \sum\limits_j a^i_j \vec{g}^{(j)}\right).
\end{equation}
Particularly accurate choice of coefficients for a time integrator corresponds to a Gauss-Legendre quadrature, where the trial directions are evaluated at the zeroes of the (shifted) Legendre polynomial
\begin{equation}
  P_n\left(2c^{(i)}-1\right) = 0,
\end{equation}
with coefficients $a^i_j$ and $b_j$ set by
\begin{eqnarray}
  \sum\limits_j a^i_j \left[c^{(j)}\right]^{k-1} &=& \frac{1}{k}\, \left[c^{(i)}\right]^k\\
  \sum\limits_j b_j \left[c^{(j)}\right]^{k-1} &=& \frac{1}{k}.
\end{eqnarray}
The resulting time integration method is A-stable and symplectic for Hamiltonian problems, and is extremely easy to implement using a simple iterative scheme.

\subsection{Static solver}\label{app:numerical:static}

Static configurations of the field $\phi$ have $\partial_t \phi = 0$ and can be found by solving a (semi-linear) elliptical problem
\begin{equation}
  {\cal L}\phi = g \Big( V'(\phi) - {\cal F} \Big).
\end{equation}
One can improve a trial solution $\bar\phi$ using Newton's method by linearizing $\phi = \bar\phi + \delta\phi$ and solving
\begin{equation}
  {\cal L}(\bar\phi + \delta\phi) = g \Big( V'(\bar\phi + \delta\phi) - {\cal F} \Big),
\end{equation}
which translates the residual ${\cal R} = -{\cal L}\bar\phi + g \Big( V'(\bar\phi) - {\cal F} \Big)$ into a correction $\delta\phi$ by solving a set of linear equations
\begin{equation}
  \Big({\cal L} - g V''(\bar\phi)\Big) \delta\phi = -{\cal L}\bar\phi + g \Big( V'(\bar\phi) - {\cal F} \Big).
\end{equation}
With the basis as chosen in the last section, this scheme converges to machine precision in about 16 iterations or so for most of the potentials.

\subsection{Inverting tortoise coordinate}\label{app:numerical:tortoise}
\label{sec:tortoise}

Accurately inverting Schwarzschild tortoise coordinate
\begin{equation}
  x = r + 2M\,\ln\left(\frac{r}{2M} - 1\right)
\end{equation}
to yield areal coordinate $r$ as a function of $x$ turns out to be a rather non-trivial task, despite appearances. The problem is that asymptotic for large positive $x$, where $r \simeq x - 2M \ln\left(x/2M - 1\right)$, and for large negative $x$, where $r \simeq 2M$ with exponentially suppressed metric function $\ln g \simeq x/2M - 1$, have vastly different derivatives with respect to $x$ (which hampers numerical schemes like Newton's method), and no closed form algebraic inverse.


A trick that works for the entire usable range of $x$ is to solve for an approximation variable $q \simeq x - 2M$ instead \begin{equation}
  q = 2M\, \ln\Bigg(\exp\left(\frac{r}{2M} - 1\right) - 1\Bigg),
\end{equation}
which (unlike $x$) is easily invertible to yield $r$
\begin{equation}
  r = 2M \Bigg(1 + \ln\left(1+\exp\frac{q}{2M}\right)\Bigg),
\end{equation}
and can be readily found by Newton's method iterating $q \rightarrow q + \delta q$ with
\begin{equation}
  \delta q = -\Bigg(r + 2M\,\ln\left(\frac{r}{2M} - 1\right) - x\Bigg) \cdot \frac{dq}{dx},
\end{equation}
as the derivative
\begin{equation}
  \frac{dq}{dx} = \left(1+\exp\frac{-q}{2M}\right) g(r)
\end{equation}
is of order one on the entire domain of definition of $x$. One still has to be careful to avoid numerical overflows in the exponents or catastrophic loss of precision when taking logarithms of one plus a small number, which can be achieved by evaluating 
\begin{equation}
  \ln\left(1+e^q\right) = \left\{ \begin{array}{rr}
    q + \ln\left(1 + e^{-q}\right),& q \ge 0\\
    2\, \text{atanh}\, {\displaystyle\frac{e^q}{2+e^q}},& q < 0\\
  \end{array} \right.
\end{equation}
in different limits.

\begin{acknowledgments}
We would like to thank Andrew DeBenedictis, Michael Desrochers, Claire Maulit, Justine Munich, Levon Pogosian, Misao Sasaki and Alexander Vikman for their valuable comments and discussions on the previous drafts of this paper. This project was partly funded by the Discovery Grants program of the Natural Sciences and Engineering Research Council of Canada and it was performed in part at the Aspen Center for Physics, which is supported by National Science Foundation grant PHY-1066293. AZ is supported in part by the Bert Henry Memorial Entrance Scholarship at SFU.
\end{acknowledgments}


\end{document}